\documentclass[aps,pra,reprint,groupedaddress,amsmath,amssymb]{revtex4-2}

\usepackage[T1]{fontenc}
\usepackage{graphicx}
\usepackage{tikz}
\usetikzlibrary{arrows.meta,positioning}
\usepackage{physics}
\usepackage{tcolorbox}
\usepackage{amsthm}
\usepackage{cancel}
\usepackage{supertabular}
\usepackage{hyperref}
\hypersetup{colorlinks=true,allcolors=blue!50!black}
\usepackage{pgfplots}
\pgfplotsset{compat=1.18}
\usepackage{cleveref}
\usepackage{placeins}
\usepackage{xcolor}

\newtheorem{theorem}{Theorem}
\newtheorem{definition}{Definition}
\newtheorem{proposition}{Proposition}
\newtheorem{lemma}{Lemma}
\newtheorem{corollary}{Corollary}
\newtheorem{remark}{Remark}

\newcommand{\good}{\textcolor{green!50!black}{\checkmark}}
\newcommand{\bad}{\textcolor{gray!55!black}{$-$}}

\newif\ifanon
\anonfalse

\begin{document}

\title{When lies look like noise: certifying untrusted entanglement swapping}

\ifanon
\author{Anonymous Authors}
\affiliation{Anonymous Institution}
\else
\author{Maria Gragera Garces}
\email{Contact author: m.gragera.garces@gmail.com}
\author{Chris Heunen}
\author{Mahesh K. Marina}
\affiliation{University of Edinburgh, Edinburgh, United Kingdom}
\fi

\date{\today}

\begin{abstract}
Entanglement swapping lets quantum repeaters distribute
entanglement over long distances, a prerequisite for the quantum
internet. We study its security in a noisy network whose
intermediate nodes perform the quantum operations faithfully but may
misreport the classical outcome that end users rely on.
We show that the ideal functionality of entanglement swapping cannot be
realised, because it gives the adversary no influence over the delivered state,
whereas a node that relabels its report does have such influence. Including 
honest noise in the ideal setting does not help unless it also concedes a
bounded fraction of misreported pairs.
We therefore propose a cut-and-choose protocol that tests a random
subset of pairs to certify the rest. 
Since no test can distinguish a misreport from readout noise, the protocol
certifies the error above a
pre-agreed minimum noise level, which is the misreport rate under our threat model. This yields a quality-of-service
guarantee, in which a failure witnesses that the node did not deliver as agreed. Unlike verification schemes that consume the
state they certify, the untested pairs remain available for use, with a
finite-size guarantee.
Certifying a $1\%$ misreport rate on $10^6$ pairs costs $10^4$ to
$1.7\times10^5$ test pairs, for minimum noise levels ranging from $0$ to $5\%$. 
The main
cost is memory, since $1\%$ requires holding $10^5$ to $10^6$ pairs at
once, far beyond the capacity of today's quantum memories.
Verifying and releasing in blocks ties the
memory to the block size at the cost of a weaker rate: a $10^4$-qubit
memory certifies $14\%$ on $4.7\times10^5$ delivered pairs.
\end{abstract}

\maketitle

\section{Introduction}
Entanglement swapping (ES) is the fundamental operation performed by quantum repeaters, enabling the distribution of entanglement over long distances~\cite{zukowski1993event,pan1998experimental,briegel1998quantum}.
It is a prerequisite for the quantum internet, whose repeater
architectures are built around it~\cite{kimble2008quantum,wehner2018quantum,azuma2023quantum}.

We investigate the security of ES in the following setting.
Two users, Alice and Bob, aim to establish a shared EPR pair.
Both have quantum computing capabilities and are connected to a wider quantum network that they do not own. All parties and channels are subject to noise. Between them sits a repeater node,
Charlie, who can, at the very least, receive and forward qubits,
perform Bell measurements, and announce classical bits.
He is trusted to perform the prescribed measurement faithfully,
but can lie about its classical
outcome.
We assume Alice, Bob, and Charlie share pairwise classical
communication channels.
The central question is:

\begin{quote}
  Can Alice and Bob establish a shared entangled pair via Charlie in a
  noisy setting, without trusting what he tells them?
\end{quote}

This scenario was not constructed for theoretical convenience. It
reflects how quantum networks are being planned and, in part, already
built. Distributing entanglement between end nodes through
intermediate repeaters is the basic service of the envisioned quantum
internet~\cite{wehner2018quantum,kozlowski2023rfc9340}, on which
applications such as distributed quantum computing (DQC)
run~\cite{cuomo2020towards,barral2025review}. The arrangement above, in
which the users supply the qubits and the relay only measures, is also how
heralded links between matter qubits work today. Each node entangles a
matter qubit with a photon and sends the photon to a midpoint station,
which interferes the incoming light and announces which detectors
clicked, and this report fixes which Bell state the two qubits
share~\cite{dahlberg2019link,pompili2021realization}. In a recent metropolitan-scale link,
the deployed fibre and the site of the midpoint station were provided by a
telecom operator, a third party to both
nodes~\cite{stolk2024metropolitan}.

Trusting the measurement but not its report reflects where the attack
surface lies. The measurement is a fixed optical set-up that can be
characterised, or self-tested~\cite{renou2018selftesting}, in advance,
whereas the report is generated by control software and carried as a
message of the network stack~\cite{dahlberg2019link,pompili2022experimental},
where a compromised or misconfigured node can alter it without touching
the optics~\cite{satoh2021attacking}. The correction it determines can be
tracked in software as a Pauli frame rather than applied
physically~\cite{knill2005quantum,riesebos2017pauli}, so altering the
record alone corrupts the pair~\cite{satoh2021attacking}. A misreport is also the cheapest deviation
available to a relay, since it costs no quantum resource, and we show
below that it cannot be told apart from noise. This trust model is the
reverse of device independence, which treats the quantum devices as
untrusted black boxes but trusts the classical software that controls
them~\cite{wehner2018quantum}.

The security of an untrusted intermediary is not a new problem.
Quantum repeater networks are known to be vulnerable to a hijacked
relay~\cite{satoh2018network}, including one that
falsifies the measurement outcomes it reports~\cite{satoh2021attacking}.
Yet the case we study, a node that preserves the quantum state but
misreports its classical messages, is explicitly left out of scope by
work on entanglement verification with tampered
nodes~\cite{amoretti2020entanglement}.

Closest to our setting, untrusted relays are handled by testing a
random subset of the distributed
entanglement~\cite{lo1999unconditional,shor2000simple,aschauer2002private},
by measurement-device-independent
QKD (MDI-QKD)~\cite{lo2012measurement,curty2014finite}, and by checking a
mediator's reported Bell outcome against a user-defined reference~\cite{krawec2015mediated,liu2018mediated,sohail2026byzantine}.
In all of these the output is a key, the analysis is asymptotic, or
the check decides only whether to abort (\Cref{ssec:related}).

Our main question is narrower. Suppose Charlie performs the swap faithfully
but may lie about its outcome. Of the pairs Alice and Bob go on to use,
on how many did he lie? And how many pairs must they test, and how many
hold in memory, to bound that number when honest hardware is itself
noisy?

Statistically, our test is the parameter-estimation step of
entanglement-based quantum key distribution (QKD), and the tools we use
are
standard~\cite{hoeffding1963probability,serfling1974probability,bouman2010sampling,tomamichel2017largely}.
The setting, however, differs from QKD in three ways
(\Cref{sssec:qkd-pe}).
First, QKD treats every error as the adversary's, while we want the error above
an honest minimum noise level.
Second, QKD feeds the bound into privacy amplification on classical data, while
our pairs pass to an application, so the guarantee must be a statement
about a quantum state.
Third, QKD users keep nothing quantum, since they measure each signal on
arrival or, in MDI-QKD, prepare and send it. 
In our situation the pairs must stay quantum, so choosing the test after Charlie's announcement requires quantum memory instead of classical memory.

Asking how often Charlie lied also turns out to have a subtle answer.
A misreport looks exactly like readout noise in Charlie's own Bell
measurement, whatever states Alice and Bob send
(\Cref{prop:unidentifiable}), so no protocol can tell a lie from a noisy
measurement. What can be certified is the error above a minimum noise level that
the users and the provider agree in advance. Under our threat model
this excess is the misreport rate, in which terms our results are stated.
Without the threat model, the same test still bounds the total
error of the batch against the agreed minimum noise level (\Cref{prop:fidelity}).

Our results have three consequences for the quantum internet. First, a
relay can be held to account by agreement but not by forensics. No protocol
can attribute an error to a lie, so the minimum noise level must be part of
the service agreement between users and provider, and our protocol makes
that agreement checkable. Second, whether a relay can be checked at all
depends on the architecture. It can when the end users prepare the pairs
and the relay only measures, as in heralded midpoint links, but not when
the relay is also the source, unless the users inject reference values of
their own (\Cref{sec:es}). Third, checking at a rate that fault-tolerant
applications can absorb requires holding $10^5$ to $10^6$ pairs at once,
so until quantum memories reach that scale, networks built from
third-party relays will have to trust their reports
(\Cref{ssec:implications}).

\subsection*{Contributions}

The following are our main results.
\begin{itemize}
\item Two impossibility results. No protocol can guarantee that Charlie
never lies. The noiseless ideal functionality $\mathcal{F}_0$ has no
composably secure realisation, and adding honest noise to it does not
help unless a non-zero misreport rate is also allowed (\Cref{thm:nogo},
\Cref{prop:batched,prop:f0-noisy}; cf.~\cite{lo1999unconditional}). Nor
can any protocol tell a lie from noise. A misreport is indistinguishable
from readout noise in Charlie's Bell measurement, whatever states Alice
and Bob send, so the minimum noise level cannot be estimated from data
and must be agreed in advance (\Cref{prop:unidentifiable}).

\item A cut-and-choose protocol in which Alice and Bob test a random
subset of the pairs in two bases and certify that Charlie lied on at
most a chosen fraction of the rest, except with a chosen probability
(\Cref{thm:security}). Variance-aware sampling bounds cut the number of
test pairs by more than an order of magnitude
(\Cref{lem:conc,lem:pooled}), and acceptance also bounds the total
error against the agreed minimum noise level (\Cref{prop:fidelity}).

\item Composable security~\cite{maurer2011abstract,portmann2022security} for the protocol (\Cref{thm:composable}),
including a bound on the quality of the final pairs when it is followed
by entanglement distillation (\Cref{prop:distil-compose}).

\item A resource estimate showing that quantum memory is the
bottleneck, and a block-wise variant that releases pairs once they are
verified, so that memory scales with the block size at the cost of a
weaker certified rate (\Cref{ssec:lifetime,ssec:blocks}).
\end{itemize}

The rest of the paper is structured as follows.
\Cref{sec:back} sets the background.
\Cref{sec:es} introduces the untrusted-relay setup, the threat model
and the ideal functionality $\mathcal{F}_0$. 
\Cref{sec:nogo} shows that
$\mathcal{F}_0$ admits no composably secure realisation, with or
without sacrificed test pairs, and that including honest noise
does not help unless a non-zero misreport rate is allowed.
\Cref{sec:protocol} gives a
cut-and-choose protocol and proves a bound on the adversary's
advantage.
\Cref{sec:composable} shows that it realises a relaxed
functionality composably and composes it with entanglement
distillation.
\Cref{sec:resources} counts the protocol's costs in test pairs and quantum memory,
and shows how block-wise verification eases the memory demand.
Finally, \Cref{sec:discussion} discusses the results and their
implications for the quantum internet.
The notation used throughout the paper is indexed in
\Cref{tab:notation} (Appendix~\ref{app:notation}) for convenience.

\section{Background} \label{sec:back}

This section sets the background and context for our work.
First, we overview the basics of entanglement swapping. 
We then review how untrusted intermediaries are
handled in classical secure computation, 
quantum repeater networks,
and other quantum communication primitives. 
Finally, we introduce cut-and-choose verification and entanglement distillation, the two tools our protocol is built upon.

\subsection{Entanglement swapping}
In its ideal, noise-free form,
ES involves two end users, Alice and Bob, and an intermediate repeater
node, Charlie.
Charlie generates two EPR pairs, $\ket{\phi^+}_{AC_1}$ and $\ket{\phi^+}_{BC_2}$,
distributing qubit $A$ to Alice and qubit $B$ to Bob while retaining $C_1$ and $C_2$.
He then performs a joint Bell measurement on $C_1$ and $C_2$ and broadcasts the two-bit classical outcome,
from which Alice and Bob apply the corresponding Pauli corrections to recover $\ket{\phi^+}_{AB}$.
In deployed architectures the two links are usually established by
heralded photonic entanglement between matter qubits held at the nodes,
with the source placed at an end node or at a midpoint
station~\cite{pompili2021realization}.
Swapping decreases the distance qubits must travel, avoiding photon loss
and decoherence that scale exponentially with
distance~\cite{briegel1998quantum}. The resulting EPR pairs are a
necessary resource for non-local gates~\cite{eisert2000optimal}, the
building blocks of DQC~\cite{barral2025review}.

\subsection{Related work}\label{ssec:related}
Our setting asks three things of a protocol: 
1) that it treat the relay's
classical report,
2) that it tolerate a minimum noise level from which deviation cannot be
distinguished, 
and 3) that the certified pairs survive for later use. 
We now discuss how classical security deals with untrusted
intermediaries, and why its methods do not carry over to the quantum setting. 
We then review
attacks on quantum repeater networks, and the security of other quantum
communication primitives, which we compare against these three
requirements in \Cref{tab:related}. 
This section ends by comparing our test to parameter estimation in QKD, which it closely resembles.

\subsubsection{Classical secure computation and fault tolerance}

Classical cryptography handles an untrusted intermediary in three main
ways: by eliminating it, by cross-checking it, or by deterring it.
\begin{itemize}
\item \emph{Elimination.} Secure multi-party computation (MPC) lets
mutually distrusting parties jointly compute a function of their
private inputs while learning nothing beyond its
output~\cite{yao1986generate,goldreich1987play}. Security is defined
against an ideal execution in which a trusted party computes the
function, and a protocol is secure if it emulates that party: the
trusted intermediary is eliminated, not verified.

\item \emph{Cross-checking.} Byzantine
agreement~\cite{lamport1982byzantine} reaches consensus when some
participants send arbitrary and inconsistent messages. It solves this by
redundancy: honest parties cross-check each other's reports.

\item \emph{Deterrence.} In the covert-adversary
model~\cite{aumann2010covert}, a party may deviate arbitrarily but is
caught with a guaranteed probability, the deterrence factor. This is
closest in spirit to our setting: our acceptance test plays an
analogous role, except that deviation must be distinguished from a
minimum noise level.
\end{itemize}

All three approaches rely on the intermediary's work being classical, so that it
can be emulated by the other parties, replicated, or re-executed for
checking. In the quantum setting, Charlie holds qubits $C_1$ and $C_2$
and is the only party who can perform the Bell measurement on them.
Since measurement is destructive and states cannot be
cloned~\cite{wootters1982single,dieks1982communication}, his role
cannot be distributed among Alice and Bob, replicated, or replayed for
verification, and checking a delivered pair consumes it. Classical MPC
results therefore do not transfer directly.

\subsubsection{Security in quantum networks}

Attacks on quantum repeaters can target confidentiality, integrity, and availability, with the
latter two introducing vulnerabilities unique to the physical constraints
of quantum communication~\cite{satoh2021attacking}.
A single hijacked repeater
can subvert naive integrity checks and bring down an entire
network~\cite{satoh2018network}, so detection mechanisms
must themselves be carefully designed.

Several works address the security of quantum communication primitives in
the presence of untrusted parties, including self-testing and
device-independent certification~\cite{fu2022parallel,coladangelo2016parallel},
quantum authentication~\cite{barnum2002authentication},
MDI-QKD and entanglement
witnesses~\cite{lo2012measurement,branciard2013mdiew}, self-testing of the
Bell-state measurement itself~\cite{renou2018selftesting,supic2020selftesting},
network nonlocality and bilocality
tests~\cite{branciard2010characterizing,branciard2012bilocal}, verifiable
blind quantum computation~\cite{fitzsimons2017unconditionally,gheorghiu2019verification},
multipartite entanglement and graph-state
verification~\cite{pappa2012multipartite,unnikrishnan2022verification}, and
efficient verification of pure quantum
states~\cite{zhu2019efficient,zhu2019general}. All assume an adversary
capable of arbitrary quantum operations, and aim to defend the resulting
resource (which is a quantum state or derived classical key) against it.
\Cref{tab:related} records where each falls short of the three
requirements our setting imposes
(and we discuss these further in Appendix~\ref{app:related-work}). 
Most share one limitation:
verification consumes the quantum resource being secured.

\begin{table*}[t]
\caption{Surveyed protocols against the three requirements of our
setting. \emph{Untrusted report}: the object of suspicion is an
intermediary's classical announcement about a measurement that cannot
be repeated. \emph{Noise tolerated}: the protocol is analysed against a
minimum noise level. \emph{Pairs survive}: the certified entanglement is still
available afterwards. 
}\label{tab:related}
\begin{ruledtabular}
\begin{tabular}{lccc}
 & Untrusted & Noise & Pairs \\
 & report & tolerated & survive \\
\hline
Self-testing~\cite{fu2022parallel}
  & \bad & \good & \bad \\
Device-independent testing~\cite{coladangelo2016parallel}
  & \bad & \good & \bad \\
Quantum authentication~\cite{barnum2002authentication}
  & \bad & \bad & \good \\
Verifiable blind computation~\cite{fitzsimons2017unconditionally}
  & \bad & \good & \bad \\
Entanglement verification~\cite{pappa2012multipartite}
  & \bad & \bad & \bad \\
Graph-state verification~\cite{unnikrishnan2022verification}
  & \bad & \good & \bad \\
Pure-state verification~\cite{zhu2019efficient}
  & \bad & \bad & \good \\
Serfling-based state verification~\cite{takeuchi2019resource}
  & \bad & \good & \good \\
Purification-based QKD~\cite{lo1999unconditional,shor2000simple}
  & \bad & \good & \good \\
Private entanglement via repeaters~\cite{aschauer2002private}
  & \bad & \good & \good \\
MDI-QKD~\cite{lo2012measurement,curty2014finite}
  & \good & \good & \bad \\
Mediated semi-quantum QKD~\cite{krawec2015mediated}
  & \good & \good & \bad \\
Mediated SQKD, fake-report analysis~\cite{liu2018mediated}
  & \good & \bad & \bad \\
Misreport traps in mediated CKA~\cite{sohail2026byzantine}
  & \good & \good & \bad \\
MDI entanglement witnesses~\cite{branciard2013mdiew}
  & \good & \good & \bad \\
Bell-measurement self-testing~\cite{renou2018selftesting}
  & \bad & \good & \bad \\
Bilocality tests~\cite{branciard2012bilocal}
  & \bad & \good & \bad \\
\hline
This work
  & \good & \good & \good \\

\end{tabular}
\end{ruledtabular}
\end{table*}

\subsubsection{Finite-size parameter estimation}\label{sssec:qkd-pe}

In entanglement-based QKD, Alice and Bob measure a random subset of
their pairs, estimate the error rate in one or two bases, and use a
sampling bound for drawing without replacement to bound the error rate
on the rest~\cite{shor2000simple,bouman2010sampling,tomamichel2012tight,tomamichel2017largely}.
Finite-key MDI-QKD likewise fixes the test rounds only after the
relay's announcement~\cite{curty2014finite} (as our operational assumption (A3) in \cref{ssec:assumptions}). Our test has
the same structure, and \Cref{lem:conc,lem:pooled} are variants of these
tools for which we claim no novelty. The setting differs in three ways,
each of which changes the result.

\begin{itemize}
\item \emph{What is bounded.} QKD attributes every error to the adversary,
so it needs an upper bound on the total error rate of the remainder and
has no use for an honest baseline. Our situation does call for a non-zero honest baseline, and the quantity of interest is the excess over it. The
guarantee therefore depends on how precisely the baseline is known
(\Cref{lem:misspec}).
Moreover, the users cannot learn this baseline from their data, because a misreport is indistinguishable from readout noise in the relay's own measurement (\Cref{prop:unidentifiable}).
\item \emph{What is kept.} In QKD the remainder is measured immediately and
only classical data is processed further; the bound feeds privacy
amplification. Here the remainder stays quantum and passes to an
application, so the guarantee must be a statement about a quantum state
(\Cref{cor:state} and \Cref{thm:composable}), and it composes with distillation
where QKD composes with hashing.
\item \emph{What it costs.} Because QKD users keep no quantum data
(measuring each signal on arrival or, in MDI-QKD, preparing and sending
it), choosing the test rounds after the announcement takes only
classical memory. Keeping the pairs instead requires quantum
memory. This binding constraint (of \Cref{sec:resources}) has no counterpart in QKD.
\end{itemize}

Purification-based proofs~\cite{lo1999unconditional,aschauer2002private}
do keep pairs, against a stronger adversary than ours, but only
asymptotically and under guarantees on the distilled output
(see Appendix~\ref{app:related-work}).

Interspersed test rounds are also already part of quantum network
design. The link layer of Dahlberg et al.\ mixes them in among the pairs
it generates, at positions both nodes fix in advance, and infers the
fidelity of the untested pairs from a finite-size sampling
bound~\cite{dahlberg2019link,tomamichel2012tight}. The heralding station
is treated there as honest. Positions fixed in advance suffice against
noise, but a station that misreports and learns them could confine its
lies to the untested pairs. This is why our test set is drawn only after
every report is in (A3), or else kept secret from the relay
(\Cref{ssec:memory}).

\subsubsection{Limits of cut-and-choose verification}\label{sssec:cutchoose-limits}

Wiesner et al.~\cite{wiesner2025cutandchoose} establish two limitations on
cut-and-choose verification of a state supplied by an untrusted source.
Both are relevant here as our protocol is of this form, although neither applies directly.

The first limitation is a no-go result: against an ideal resource that, like
$\mathcal{F}_0$ below, returns either the target state or an abort, all
a simulator can do is choose an abort probability, and no protocol is
composably secure. Their bound,
$\varepsilon_H + \varepsilon_D \geq 1/4\sqrt{N}$ over $N$ verification
rounds, holds in universal composability, abstract cryptography and
categorical composable cryptography~\cite{broadbent2023categorical}
alike. \Cref{thm:nogo} has the same shape, but does not decay in $N$.
In their setting the clients measure copies of the state they keep, so
any significant deviation shows up in the sample. Here the report and
the pair are different objects: the attack of \Cref{lem:attack} shifts a
delivered pair onto an orthogonal Bell state, and only a measurement of
that pair would detect it. The attack leaves no trace on the tested
pairs at all, so the gap is $1-q$ rather than $1/4\sqrt{N}$.

The second limitation is a trade-off between efficiency and security. 
When the untrusted
source emits $N+1$ registers and the clients measure $N$ and keep one,
the correctness error $\varepsilon_H$ and security error
$\varepsilon_D$ satisfy $\varepsilon_H + \varepsilon_D \geq 1/7N$ for a
fidelity-based standalone notion and $\geq 1/4\sqrt{N}$ for any
composable one. This bound does not apply to our protocol, which
differs in four respects. 
First, Alice and Bob prepare their own pairs, so the suspected party
emits nothing. 
Second, we keep $M$ pairs and test $k$, with $M$ some six times
$k$ at the operating point of \Cref{ssec:test-overhead}. 
Third, our security parameter is a fraction of the
delivered pairs rather than the fidelity of the kept state to a pure
target, so one corrupted pair in $10^6$ is not a large error. 
Fourth, the kept pairs are not returned untouched: Alice and Bob apply the
correction $P_m$, which is the operation being tested.

\subsubsection{Entanglement distillation}\label{sssec:distillation}

Entanglement
distillation~\cite{bennett1996purification,deutsch1996quantum} has
Alice and Bob consume several noisy pairs, using local operations
and classical communication, to produce fewer pairs
of higher fidelity. It is the standard way of handling noise in
distributed quantum computation~\cite{cirac1999distributed}. 
No intermediary participates. Distillation does not replace
verification, since it addresses a different problem: it improves
fidelity conditional on a bounded input error rate, whereas a security
protocol certifies that bound against an adversary who controls both the
errors and the reports describing them. Verification is therefore a
precondition for applying distillation with a guarantee.

\section{Secure entanglement swapping}\label{sec:es}

\subsection{The failures of standard ES}

In the standard ES architecture~\cite{zukowski1993event,wehner2018quantum},
Charlie generates both EPR pairs, a choice motivated by the need for
simultaneous, spectrally identical photons~\cite{zukowski1993event}, a
hardware requirement since met with independent
sources~\cite{halder2007entangling}.
This poses the security problem that Charlie
controls the entire preparation: he may retain a purification or
prepare something other than $\ket{\phi^+}$, while arranging the state
to explain whatever statistics Alice and Bob receive.
Unless the users inject values of their own, as in mediated
semi-quantum key distribution~\cite{krawec2015mediated,liu2018mediated},
where each user measures and resends some qubits and checks the
server's Bell report against what they resent, no quantity in the
execution has a value they know independently of him, so there is
nothing against which a deviation can be measured.

\subsection{Untrusted-relay ES setup}

The setup used throughout this work instead reassigns the source role to the end users:

\begin{enumerate}
\item Alice generates $\ket{\phi^+}_{A_1 A_2}$ and Bob generates
$\ket{\phi^+}_{B_1 B_2}$.
\item Alice sends $A_2$ and Bob sends $B_2$ to Charlie; they retain
$A_1$ and $B_1$.
\item Charlie performs the Bell measurement
  on $A_2$ and $B_2$, obtaining outcome $r \in \{0,1\}^2$, and
  announces a report $m$. An honest Charlie sets $m = r$; we keep the two symbols distinct because the difference between them is what we study.
\end{enumerate}

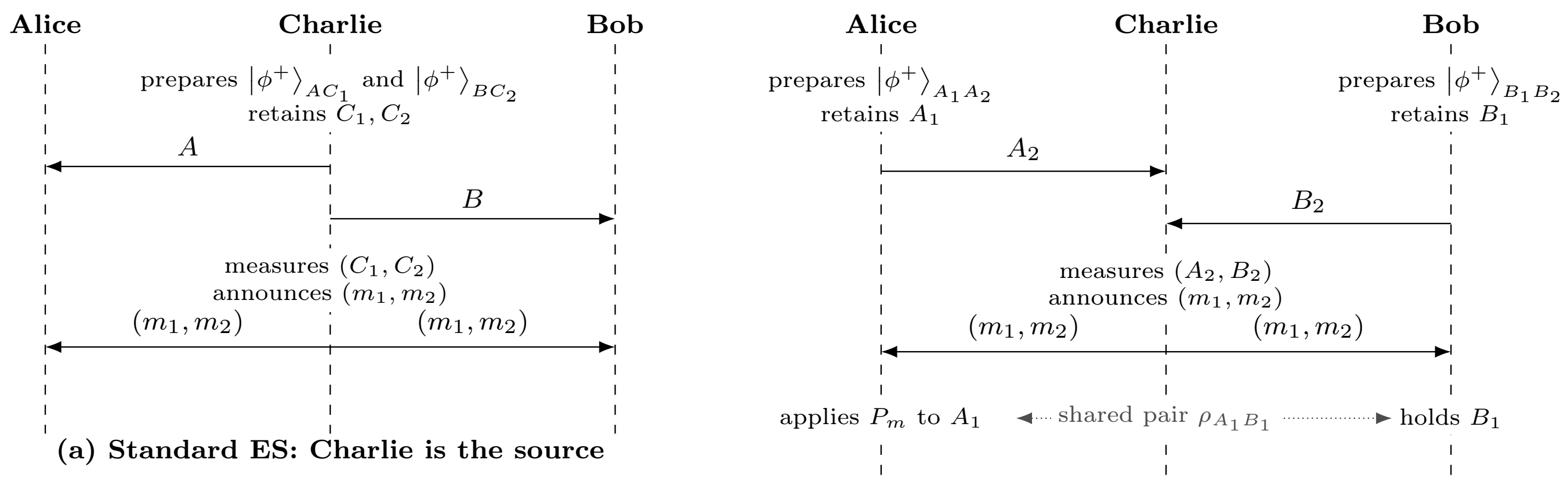
\begin{figure*}[t]
\centering
\resizebox{\textwidth}{!}{%
\begin{tikzpicture}[
  >=Latex,
  every node/.style={font=\footnotesize},
  actor/.style={font=\bfseries\footnotesize},
  note/.style={font=\scriptsize, align=center, inner sep=2pt, fill=white}
]

\node[actor] (A1) at (0,0) {Alice};
\node[actor] (C1) at (3,0) {Charlie};
\node[actor] (B1) at (6,0) {Bob};
\draw[dashed] (A1.south) -- ++(0,-4.1);
\draw[dashed] (C1.south) -- ++(0,-4.1);
\draw[dashed] (B1.south) -- ++(0,-4.1);

\node[note] at (3,-0.75)
  {prepares $\ket{\phi^+}_{AC_1}$ and $\ket{\phi^+}_{BC_2}$\\ retains $C_1, C_2$};
\draw[->] (3,-1.5) -- node[above]{$A$} (0,-1.5);
\draw[->] (3,-2.05) -- node[above]{$B$} (6,-2.05);
\node[note] at (3,-2.7) {measures $(C_1,C_2)$\\ announces $(m_1,m_2)$};
\draw[->] (3,-3.4) -- node[above]{$(m_1,m_2)$} (0,-3.4);
\draw[->] (3,-3.4) -- node[above]{$(m_1,m_2)$} (6,-3.4);
\node at (3,-4.5) {\textbf{(a) Standard ES: Charlie is the source}};

\begin{scope}[xshift=8.8cm]
\node[actor] (A2) at (0,0) {Alice};
\node[actor] (C2) at (3,0) {Charlie};
\node[actor] (B2) at (6,0) {Bob};
\draw[dashed] (A2.south) -- ++(0,-4.6);
\draw[dashed] (C2.south) -- ++(0,-4.6);
\draw[dashed] (B2.south) -- ++(0,-4.6);

\node[note] at (0,-0.75) {prepares $\ket{\phi^+}_{A_1A_2}$\\ retains $A_1$};
\node[note] at (6,-0.75) {prepares $\ket{\phi^+}_{B_1B_2}$\\ retains $B_1$};
\draw[->] (0,-1.55) -- node[above]{$A_2$} (3,-1.55);
\draw[->] (6,-2.1) -- node[above]{$B_2$} (3,-2.1);
\node[note] at (3,-2.75) {measures $(A_2,B_2)$\\ announces $(m_1,m_2)$};
\draw[->] (3,-3.45) -- node[above]{$(m_1,m_2)$} (0,-3.45);
\draw[->] (3,-3.45) -- node[above]{$(m_1,m_2)$} (6,-3.45);
\node[note] at (0,-4.15) {applies $P_{m}$ to $A_1$};
\node[note] at (6,-4.15) {holds $B_1$};
\draw[<->, densely dotted, gray!60!black] (1.42,-4.15) -- (5.38,-4.15);
\node[font=\scriptsize, gray!60!black, fill=white, inner sep=2pt]
  at (3,-4.15) {shared pair $\rho_{A_1B_1}$};
\node at (3,-5.0) {\textbf{(b) Untrusted-relay ES: end users are the sources}};
\end{scope}

\end{tikzpicture}}
\caption{Message flow in (a) standard ES, where Charlie generates both
EPR pairs, and (b) the untrusted-relay setup, where Alice and Bob
generate their own pairs and Charlie only measures and reports.}
\label{fig:es-comparison}
\end{figure*}

The setup adds no communication overhead: each user sends one qubit and
Charlie announces one outcome, as before. The only change is that Charlie
no longer controls the preparation. Alice and Bob jointly know their states
before the swap, so for any announced outcome they can predict
the post-correction state and compare it against their measurements. Our
protocol leverages this comparison against a reference fixed by their own preparation.

\subsection{Threat model}\label{ssec:threat-scope}

Charlie receives one qubit from each party, performs the prescribed Bell
measurement, and announces two classical bits. The correction $P_m$ is
the only use either party makes of the report, so it is the only route
a misreport can be observed. The threat we address is
selective classical misreporting: Charlie performs every quantum
operation faithfully but may lie about the outcome, on any chosen pairs. This is a narrower adversary than the arbitrary quantum
deviation assumed by the works of \Cref{sec:back}: we trust the quantum
operation and withhold trust only from the classical report of its
outcome. We thus certify the part that self-testing of the Bell
measurement~\cite{renou2018selftesting} does not cover; running such a
test before our protocol would cover both.

Our threat model excludes two cases. First, Charlie
retains no side information about the delivered pair. We do not
consider purification attacks, in which Charlie measures non-projectively
and keeps auxiliary qubits entangled with the delivered ones. The statistics of a
single noisy pair cannot rule out such an attack, since the partial
trace over any auxiliary system is many-to-one~\cite{HJW1993}. Such
correlations can be removed, by entanglement distillation after a
fidelity test~\cite{deutsch1996quantum,lo1999unconditional,aschauer2002private}
or certified away device-independently~\cite{coladangelo2016parallel,fu2022parallel},
but both consume pairs this protocol delivers, and we do not analyse
them here. Second, a Charlie who measures something other than the
prescribed Bell basis has no outcome $r$ to misreport, so the misreport
rate $f$ of \Cref{def:prob-security} is not defined.

We make three further assumptions about the setting. The operational
assumptions (A1)--(A7) of the protocol are described in \Cref{ssec:assumptions}.
\begin{enumerate}
\item[(T1)] The transit noise between the users and Charlie is
unital, so that his outcome $r$ is uniform (\Cref{lem:attack}); under
non-unital noise such as amplitude damping, the distribution of honest
reports is biased, and a relabelled report shifts it, which gives the
users an additional check we do not exploit.
\item[(T2)] Charlie's report of whether a round succeeded is trusted; only
the outcome he reports for a successful round is not. Success is
reported separately because a linear-optical Bell measurement resolves
at most two of the four Bell states~\cite{calsamiglia2001maximum}, and
heralded architectures announce it as a second classical
message~\cite{pompili2021realization}. If this assumption fails, a
falsely heralded round has no outcome to misreport and lies outside
$f$, but \Cref{prop:fidelity} still bounds its effect on the fidelity
delivered.
\item[(T3)] The users' own preparation and measurement are trusted;
(A4) states what they must be able to do.
\end{enumerate}

Not every result in this paper depends on these exclusions though.
\Cref{prop:fidelity} holds against an arbitrary instrument applied pair
by pair, so the fidelity of the delivered pairs is bounded whether or
not Charlie measured as prescribed or kept a purification. 
Nonetheless, no measurement on $A_1$ and $B_1$ alone can bound
the information Charlie kept.

\subsection{The ideal functionality}

To define security we compare the protocol against an ideal
functionality, which a secure
protocol should perform indistinguishably from.

\begin{definition}[Noiseless ES functionality $\mathcal{F}_0$]\label{def:ideal-F}
The functionality $\mathcal{F}_0$ interacts with Alice, Bob and an
adversary, Charlie, as follows.
\begin{enumerate}
\item Alice submits $A_2$ and Bob submits $B_2$; they retain $A_1$ and
$B_1$.
\item $\mathcal{F}_0$ discards $A_2$ and $B_2$, draws
$m = (m_1,m_2) \in \{0,1\}^2$ uniformly at random, and announces $m$ to
all three parties.
\item Charlie sends a bit: if it is $0$, then $\mathcal{F}_0$ sets Alice and
Bob's registers $A_1B_1$ to $\ket{\phi^+}$; if it is $1$, then $\mathcal{F}_0$ outputs $\bot$ to
all three parties.
\end{enumerate}
\end{definition}

The announcement in step~2 signifies that an 
honest Charlie's outcome is uniformly random and carries no
information. Charlie can therefore stop a delivery, but cannot
influence the delivered state. 

Further, $\mathcal{F}_0$ has no fidelity
parameter because it delivers exact states; tolerating noise requires
a relaxed functionality (\Cref{def:F-relaxed}). 
We write $\delta$ for the infidelity of states in a real execution.

\begin{remark}\label{rem:trivial-nogo}
If an honest Charlie delivers $\rho_{AB}$ with
$F(\rho_{AB},\ket{\phi^+}\bra{\phi^+}) = 1-\delta$, then an environment that
outputs $1$ if the execution aborted, and otherwise performs a Bell
measurement on the delivered pair and outputs $1$ on outcome
$\ket{\phi^+}$, accepts with probability $1-\delta$ in the real
execution and with probability $1$ in the ideal one, whichever branch
the simulator takes. The composable security error is thus
$\eta \geq \delta$, with or without an adversary.
\end{remark}

This trivial bound vanishes as $\delta \to 0$, whereas the gap shown in
\Cref{sec:nogo} does not.

\section{No-go for composable security} \label{sec:nogo}

A real execution has a channel that $\mathcal{F}_0$ does not, so the
delivered state $\rho_{AB}$ has fidelity at least $1-\delta$ with
$\ket{\phi^+}$ for a noise tolerance $\delta$.
This section measures the resulting gap.

That $\mathcal{F}_0$ cannot be realised is not in itself surprising: it
gives the adversary no influence over the delivered state at all. The
question is what must be conceded to repair it. We show that without
noise a single undetected lie already breaks it, and, at the end of the
section, that an ideal object which includes honest noise still
fails unless it also allows for a non-zero misreport rate.

\begin{definition}[Environment]\label{def:environment}
An \emph{environment} $\mathcal{E}$ is an arbitrary
(computationally unbounded) algorithm that may supply inputs
to Alice and Bob, observe all their outputs, including the delivered
quantum registers on which it may perform arbitrary joint
measurements, and use those outputs in arbitrary
computations later; it outputs a single bit.
\end{definition}

\begin{definition}[Simulator]\label{def:simulator}
A \emph{simulator} $S$ is an algorithm that, given
access to an ideal functionality $\mathcal{F}$ and to the
inputs and outputs of the honest parties, attempts to reproduce
the view that a real dishonest Charlie $C^*$ would generate.
We grant $S$ every capability the standard
formulations allow, as well as some more:
$S$ is computationally unbounded, has black-box access to $C^*$,
and may instruct $\mathcal{F}$ to abort.
\end{definition}

A no-go result proved against the strongest simulator holds against every
weaker one. Since the environment may measure a
delivered quantum register, classical universal
composability~\cite{canetti2020uc} does not suffice, and abstract
cryptography~\cite{maurer2011abstract} deliberately leaves the model of
systems open. We
specialise the quantum treatments of Unruh~\cite{unruh2010universally}
and of Portmann and Renner~\cite{portmann2022security}, the latter an
instance of abstract cryptography, to a single
corrupted party and a single functionality, in the manner of
Dunjko et al.~\cite{dunjko2014composable}.

\begin{definition}[Composable security]\label{def:composable}
Let $\mathcal{F}$ be an ideal functionality. A protocol $\Pi$ is said to
\emph{realise $\mathcal{F}$ with composable security error}
$\eta \in [0,1]$ if for every dishonest Charlie $C^*$ there
exists a simulator $S$ such that for
\emph{every} environment $\mathcal{E}$:
\begin{equation}
\left| \Pr[\mathcal{E}(\text{real}_{\Pi, C^*}) = 1]
  - \Pr[\mathcal{E}(\text{ideal}_{\mathcal{F}, S}) = 1] \right| \leq \eta
\end{equation}
where $\text{real}_{\Pi, C^*}$ denotes the output of a real execution
of $\Pi$ against $C^*$, and
$\text{ideal}_{\mathcal{F}, S}$ denotes the output of $\mathcal{F}$
with simulator $S$. The protocol achieves
\emph{perfect} composable security when $\eta = 0$.
\end{definition}

\Cref{def:composable} is a statement about distinguishing real and ideal situations;
that a protocol satisfying it may be substituted for $\mathcal{F}$
inside a larger one, with errors adding, is a property of the frameworks
cited~\cite{unruh2010universally,portmann2022security}. On its own it
is trivially satisfiable: the protocol that always aborts realises
$\mathcal{F}_0$ with $\eta = 0$. A no-go result must therefore require the
protocol to deliver something.

\begin{definition}[Completeness]\label{def:completeness}
A protocol $\Pi$ is \emph{$\mu$-complete} if it aborts with probability
at most $\mu$ on an honest execution, that is, when Alice, Bob and
Charlie all follow the specification.
\end{definition}

Completeness rules out this trivial solution, as $\mu$ bounds how often
the protocol may decline to deliver. A protocol with the
interface of $\mathcal{F}_0$ receives one qubit from Alice and one from
Bob and returns one pair, so any check it performs acts either on the
qubits in transit or on the pair it is about to deliver.

\begin{lemma} \label{lem:attack}
Let $\Pi$ be an entanglement swapping protocol with the interface of
$\mathcal{F}_0$, and write $\rho^{\mathrm{hon}}_{AB}$ for the state it
delivers on an honest execution. Charlie has a dishonest strategy $C^*$
such that:
\begin{enumerate}
  \item the classical report $(m_1, m_2)$ of $C^*$ is identically
  distributed to that of an honest execution;
  \item everything on which $\Pi$ can base its accept-or-abort decision
  is distributed exactly as in an honest execution, so if $\Pi$ is
  $\mu$-complete then it aborts against $C^*$ with probability at
  most $\mu$;
  \item whenever $\Pi$ accepts, the state it delivers is
  \begin{equation}\label{eq:spoofedstate}
    \rho_{AB} = (P_e \otimes I)\,\rho^{\mathrm{hon}}_{AB}\,
    (P_e \otimes I)^\dagger,
  \end{equation}
  where Charlie chooses
  $P_e \in \{\sigma_x, \sigma_z, \sigma_x\sigma_z\}$ beforehand at
  will. In the noiseless case
  $\rho^{\mathrm{hon}}_{AB} = \ket{\phi^+}\bra{\phi^+}$ this is a Bell
  state orthogonal to the target.
\end{enumerate}
\end{lemma}
\begin{proof}
Let Charlie obtain Bell outcome $r \in \{00,01,10,11\}$, on which Alice
and Bob apply the corresponding Pauli correction $P_r$. Define $C^*$ as
follows:
\begin{enumerate}
\item Perform the Bell measurement honestly to obtain $r$; perform
all required auxiliary qubit measurements honestly.
\item Report $m = r \oplus e$ for a fixed non-zero
$e \in \{01,10,11\}$; report all auxiliary results truthfully.
\end{enumerate}

\textit{Property 1:} Under unital transit noise, which we assume
(\Cref{ssec:threat-scope}), the state of $A_2B_2$ received by Charlie
is maximally mixed, so $r$ is uniformly distributed
over $\{00,01,10,11\}$. Since $e$ is fixed, $m = r \oplus e$ is also
uniformly distributed, and the report of $C^*$ is identically
distributed to that of an honest execution.

\textit{Property 2:} Let $\mathcal{V}$ collect everything on which $\Pi$
may condition its accept-or-abort decision: all classical traffic, the
qubits in transit, and any auxiliary registers, but \emph{not} the
delivered pair, which a protocol with the interface of $\mathcal{F}_0$
must hand to Alice and Bob rather than measure. We show that the joint
state of $\mathcal{V}$ is the same under $C^*$ as under an honest
Charlie, so that no decision rule applied to
$\mathcal{V}$ can distinguish the two, and $\Pi$ aborts against
$C^*$ with the probability it aborts on an honest execution,
which is at most $\mu$ when $\Pi$ is $\mu$-complete.

$C^*$ applies no operation to any register. Charlie forwards each qubit
unaltered, performs the prescribed Bell measurement, and reports every
auxiliary result truthfully, so the quantum contents of $\mathcal{V}$
are produced by the same channel and the same measurement in both
executions. The single deviation is to output $m = r \oplus e$ instead
of $r$, and by property~1 that string is uniform on $\{0,1\}^2$ in both.
It remains to check that the announcement is not correlated with the
rest of $\mathcal{V}$ differently in the two: $r$ is
correlated with the post-measurement state of $A_1B_1$, which is
precisely the delivered pair and is excluded from $\mathcal{V}$, and
with nothing else, since the measured qubits $A_2, B_2$ are consumed in
producing it. Relabelling a uniform string that is uncorrelated with
everything in $\mathcal{V}$ leaves the joint state of $\mathcal{V}$
unchanged.

\textit{Property 3:} Consider the case where $\Pi$ accepts. Alice and Bob apply
the correction $P_m$ based on the announced $m$. Since $m = r \oplus e$, they apply $P_{r \oplus e}$
where an honest execution would have applied $P_r$. The two differ by
$P_{r \oplus e}P_r^\dagger = P_e$, up to a global phase, which is
unobservable and which we suppress throughout. The correction is the
last operation either party performs, so whatever state the swap and
the channel have left them holding, the delivered state
is~\cref{eq:spoofedstate}. This covers a noisy execution as well as a
noiseless one: $C^*$ does not touch the noise, he relabels the
correction applied on top of it.
\end{proof}

In a noisy setting this attack is indistinguishable from channel
noise~\cite{deutsch1996quantum,aschauer2002private}. 
Both produce a shifted $\rho_{AB}$ with a classical report consistent with an
honest execution, and since the local reduced states are maximally mixed
whichever Bell state $\rho_{AB}$ is, neither party can detect it
locally.

\begin{theorem}[No-go for composable security] \label{thm:nogo}
Let $\Pi$ be a $\mu$-complete entanglement swapping protocol with the
interface of
$\mathcal{F}_0$, and let $p_{\phi^+} = 1-\delta$ together with
$p_{\phi^-}, p_{\psi^+}, p_{\psi^-}$ be the Bell-diagonal weights of the
state $\rho^{\mathrm{hon}}_{AB}$ it delivers on an honest execution.
Then $\Pi$ does not achieve composable security with respect to
$\mathcal{F}_0$ with error
\begin{equation}\label{eq:eta-bound}
  \eta \;<\; 1 - \mu - q, \qquad
  q \;=\; \min_{B \,\neq\, \phi^+} p_B \;\leq\; \frac{\delta}{3} .
\end{equation}
In particular $\eta \geq 1-\mu$, which is close to $1$ for any
protocol that rarely aborts,
whenever the honest execution leaves some Bell component other
than $\ket{\phi^+}$ unpopulated. That includes the noiseless case
$\delta = 0$, and any purely dephasing channel.
\end{theorem}

\begin{proof}
Choose $e$ so that $P_e\ket{\phi^+}$ is the Bell state carrying the
least weight in $\rho^{\mathrm{hon}}_{AB}$, and run the $C^*$ of
\Cref{lem:attack} with that $e$. By property~2 the protocol accepts, and
by property~3 it delivers
$\rho_{AB} = (P_e\otimes I)\rho^{\mathrm{hon}}_{AB}(P_e\otimes I)^\dagger$,
whose weight on the target is
\begin{equation}
  \bra{\phi^+}\rho_{AB}\ket{\phi^+}
  = \bra{P_e\phi^+}\rho^{\mathrm{hon}}_{AB}\ket{P_e\phi^+}
  = q .
\end{equation}

Let $\mathcal{E}$ measure the delivered pair in the Bell basis and
output $1$ if the execution aborted, or if it did not abort and the
outcome is $\ket{\phi^+}$. By property~2 the real execution aborts with
probability at most $\mu$, and on the complementary event it delivers a
state whose Bell measurement returns $\ket{\phi^+}$ with probability
$q$, so
\begin{equation}
  \Pr\big[\mathcal{E}(\text{real}) = 1\big]
  \;\leq\; \mu + (1-\mu)\,q \;\leq\; \mu + q ,
\end{equation}
the first bound being increasing in the abort probability.
The ideal execution either aborts, in which case $\mathcal{E}$ outputs
$1$, or delivers $\ket{\phi^+}_{AB}$, in which case the Bell
measurement returns $\ket{\phi^+}$ with certainty and $\mathcal{E}$
outputs $1$ again:
\begin{equation}
  \Pr\big[\mathcal{E}(\text{ideal}) = 1\big] = 1 .
\end{equation}
This holds whatever $S$ does, and in particular however it randomises
between the two branches. Hence $\eta \geq 1-\mu-q$. The three weights $p_B$
with $B \neq \phi^+$ sum to $\delta$, so their minimum is at most
$\delta/3$, and $q = 0$ as soon as one of them vanishes.
\end{proof}

\begin{remark}\label{rem:simulator}
The proof places no constraint on $S$, so the extra capabilities of
\Cref{def:simulator} do not help. Nor does randomising between
aborting and delivering, since the environment of \Cref{thm:nogo}
outputs $1$ on both ideal branches. The obstruction is that
$\mathcal{F}_0$ gives the ideal adversary no influence over the
delivered state, which a real adversary does
(\Cref{lem:attack}).
\end{remark}

\Cref{thm:nogo} concerns protocols with the single-pair interface of
$\mathcal{F}_0$, which have no spare pairs to test. The natural
objection is that this rules out testing by construction. The remedy is
to prepare $N$ pairs, test $k$ of them and deliver the rest, as the
protocol of \Cref{sec:protocol} does. Against $\mathcal{F}_0$ this does
not help, as an adversary who misreports a single, randomly chosen pair is
missed by the test with probability $1-k/N$.

\begin{proposition}\label{prop:batched}
Let $\Pi$ be $\mu$-complete, prepare $N$ pairs, set aside any $k$ of them for testing by
whatever means it likes, and on acceptance deliver the remaining
$M = N-k$, realising $\mathcal{F}_0^{\otimes M}$; \Cref{def:composable}
is read on that product, with a single simulator facing an environment
that sees all $M$ delivered pairs. Then $\Pi$ does not
achieve composable security with error
\begin{equation}\label{eq:eta-batched}
  \eta \;<\; 1 - \mu - \frac{k}{N} - q ,
\end{equation}
with $q$ as in \Cref{thm:nogo}.
\end{proposition}
\begin{proof}
Let $C^*$ draw an index $i \in \{1,\dots,N\}$ uniformly at random, run
the strategy of \Cref{lem:attack} on pair $i$, and behave honestly on
the others. By property~1 his reports carry no information about $i$, so
the test set is independent of it and $\Pr[i \in T] = k/N$. Let
$\mathcal{E}$ output $1$ if the execution aborted, or if it did not
abort and a Bell measurement on every delivered pair returns
$\ket{\phi^+}$. As in \Cref{thm:nogo},
$\Pr[\mathcal{E}(\text{ideal})=1] = 1$. In the real execution,
$\mathcal{E}$ outputs $1$ only on one of three events. The first,
$i \in T$, has probability $k/N$. The second is that $i \notin T$ and
the execution aborts. On $i \notin T$ the tested pairs are all honest
and pair $i$ is delivered, so by property~2 the joint distribution of
this event and the decision is the same as in an honest execution, and
the second event has probability at most $\mu$. The third is that
$i \notin T$, the execution accepts, and pair $i$ returns
$\ket{\phi^+}$, which has probability at most $q$. Hence
\begin{equation}
  \Pr[\mathcal{E}(\text{real})=1] \;\leq\; \frac{k}{N} + \mu + q .
\end{equation}
\end{proof}

\begin{remark}[How to read \cref{eq:eta-batched}]\label{rem:batched-reading}
Under noise the bound says little about the adversary. An honest
execution passes the environment of the proof only with probability
$(1-\delta)^M$, negligible at the batch sizes of \Cref{sec:resources},
so $\eta \geq 1 - \mu - (1-\delta)^M$ holds with no adversary at all
(\Cref{rem:trivial-nogo}). The adversary's contribution shows at
$\delta = 0$: a single misreport evades detection with probability
$1-k/N$, so the difference shows only when $k/N \to 1$.
\end{remark}

Testing a subset therefore does not help against $\mathcal{F}_0$.

\paragraph*{Including the noise does not help.}
The natural repair is an ideal object that already delivers the honest
noise. \Cref{def:F-relaxed} in \Cref{sec:composable} is one: it applies
the honest channel to each delivered pair and lets the adversary
misreport on at most a fraction $f$ of them, so at $f = 0$ it concedes
the noise but no lies. The protocol of \Cref{sec:protocol} still does
not realise it (\Cref{prop:f0-noisy}). The composable error is at least
$0.43$ at the operating point of \Cref{ssec:test-overhead} and $0.99$ at
the loose point, and since an honest execution realises this ideal
exactly, the whole gap is due to Charlie. It comes from a difference in
resolution. The test sees only $k$ pairs and cannot detect a misreport
rate much below its margin, while an environment that measures all $M$
delivered pairs resolves far smaller shifts. A Charlie who misreports at
a rate between the two passes the test and is still seen by the
environment. Any test that measures $k \ll M$ pairs leaves this possibility
open, so a non-zero misreport rate must be conceded because of 
adversaries, not because of noise.

The alternative is to grant the ideal-world adversary the same influence as
the real one, as
MDI-QKD~\cite{lo2012measurement,curty2014finite} does: it is composably
secure with an untrusted relay performing our Bell measurement, but its
ideal object may abort at the adversary's discretion, is parameterised
by an error rate the simulator may set, and delivers a classical key.
\Cref{sec:composable} proposes an analogue for a functionality that
delivers entangled pairs instead of a key.

\section{A cut-and-choose protocol for probabilistic security} \label{sec:protocol}

Since $\mathcal{F}_0$ cannot be realised, we aim for a weaker guarantee.
Instead of requiring every delivered pair to be correct, we require
that whenever the protocol accepts, Charlie misreported on at most a
chosen fraction of the delivered pairs, except with a small probability
fixed in advance.
We define this notion of probabilistic
security, state our assumptions, and give a cut-and-choose protocol
that achieves it.
We then prove that acceptance bounds the misreport rate, show that
acceptance also
bounds the delivered fidelity against an arbitrary relay. Composability is deferred to
\Cref{sec:composable}.

\subsection{Probabilistic security}

\begin{definition}[Probabilistic security]\label{def:prob-security}
Let $f_{\mathrm{target}} \in [0,1]$ and let
$\varepsilon_{\mathrm{sec}}\geq 0$ be a security parameter. Let $\Pi$
be an entanglement swapping protocol which, on acceptance, delivers a
set $D$ of $M$ pairs to Alice and Bob. For an execution of $\Pi$,
write $r^{(i)}$ and $m^{(i)}$ for the outcome Charlie obtained on pair
$i$ and the outcome Charlie announced, and define the \emph{deviation rate
on the delivered pairs}
\begin{equation}\label{eq:f-def}
  f \;=\; \frac{\bigl|\{\, i \in D \;:\; m^{(i)} \neq r^{(i)} \,\}\bigr|}{M}.
\end{equation}
Then $\Pi$ is \emph{$(f_{\mathrm{target}}, \varepsilon_{\mathrm{sec}})$-secure} if,
for every strategy of Charlie within the threat model of
\Cref{ssec:threat-scope},
\begin{equation}\label{eq:sec-joint}
  \Pr\bigl[\,\Pi \text{ accepts} \;\wedge\; f > f_{\mathrm{target}}\,\bigr]
  \;\leq\; \varepsilon_{\mathrm{sec}} .
\end{equation}
\end{definition}

The bound is on the joint event.
A conditional statement cannot hold uniformly, since a strategy that is
almost always rejected makes the conditional probability large at no
risk to the users~\cite{lo1999unconditional,tomamichel2012tight}.

The rate $f$ is taken over the delivered pairs, since test pairs
never reach the application, and counts only misreports. 
Channel noise is accounted for by the honest baselines
$\gamma_Z, \gamma_X$, which account for the noise levels an honest execution shows when
the users measure a pair in the $Z$ and $X$ bases, agreed upon in advance. 
Only errors in excess of
this declared minimum noise level are attributed to malicious action by Charlie.

\subsection{Assumptions}\label{ssec:assumptions}

Beyond the setup and threat model of \Cref{sec:es}, including
(T1)--(T3), the analysis to follow 
requires the following assumptions.

\begin{enumerate}
\item[(A1)] \textbf{Agreement on Charlie's reports:}
Alice and Bob hold the same report for each pair. Over pairwise
channels a dishonest Charlie could send Alice and Bob different
reports for the same pair; to rule this out, they compare what they received
over (A2) and abort on any mismatch, as in mediated key
distribution~\cite{krawec2015mediated}. 
We do not assume a broadcast channel from Charlie; even if one existed,
it would only force Charlie to send both users the same report, not
necessarily a true one.

\item[(A2)] \textbf{An authenticated classical channel between Alice
and Bob:} 
is assumed and used to exchange test outcomes and compare reports. Comparing
all $N$ reports costs a single universal
hash~\cite{wegman1981new} evaluated with shared
randomness, which is information-theoretically secure provided the hash
function is drawn after all reports have been received. The channel
need only be authenticated, not secret.

\item[(A3)] \textbf{Randomness drawn after Charlie has committed:}
Alice chooses which pairs to test, in which basis, and a random order
for the rest, only after Charlie has announced all his reports. 
She then sends these choices to Bob (A2). 
Because they are chosen after Charlie can no longer change his reports,
there is no need for this communication to be secret. Charlie may
therefore learn all of them, including the delivery order; the analysis
allows for this, and the ideal functionality of \Cref{def:F-relaxed}
hands them to the adversary explicitly.

\item[(A4)] \textbf{Local state preparation and measurement:} Alice
and Bob can each prepare $\ket{\phi^+}$ locally and measure in both
the $Z$ and $X$ bases.

\item[(A5)] \textbf{Quantum memory:} Both parties can hold all $N$ of
their retained halves at once, for the duration given
in~\cref{eq:coherence}. Capacity and duration are separate demands,
and together they are the most costly assumption in practice
(\Cref{ssec:memory}).

\item[(A6)] \textbf{Agreed parameters:} $M$, $k$, $\gamma_Z$,
$\gamma_X$, $f_{\mathrm{target}}$ and $\varepsilon_{\mathrm{sec}}$ are
fixed before the protocol begins. The baselines are declared as batch
averages, for a batch of the same size held for the same duration. The
users measure the components that do not involve the relay themselves;
the relay's own share cannot be learned from protocol data
(\Cref{prop:unidentifiable}) and is a trusted input
(\Cref{rem:overest}). 

\item[(A7)] \textbf{Independent honest noise:} Given the per-pair
rates, which may differ from pair to pair (\Cref{rem:index-noise}), the
honest channel noise acts independently across pairs. Correlated noise,
such as calibration drift within a batch, would inflate the variance of
the test statistic beyond what \Cref{lem:conc} allows; a martingale
bound~\cite{azuma1967weighted} would relax this, and \Cref{lem:pooled} does so for noise
correlated between blocks, though not within one.
\end{enumerate}

\subsection{Protocol}\label{ssec:protocol}

The protocol is a cut-and-choose
test~\cite{lindell2007efficient,bouman2010sampling} split between two
bases (similarly to the check step of purification-based
QKD~\cite{shor2000simple}). 
It takes as input the parameters agreed in (A6): the
number $M$ of pairs to deliver, the number $k$ of test pairs (even, so
that they split equally between the two bases), the baselines
$\gamma_Z, \gamma_X$, the target $f_{\mathrm{target}}$, and the security
parameter $\varepsilon_{\mathrm{sec}}$. It runs as follows.

\begin{enumerate}
  \item \textbf{Preparation:} Alice generates $\ket{\phi^+}_{A_1A_2}$,
  retains $A_1$, and sends $A_2$ to Charlie. Bob generates
  $\ket{\phi^+}_{B_1B_2}$, retains $B_1$, and sends $B_2$ to
  Charlie. This is repeated $N = M + k$ times.
  \item \textbf{Measurement and announcement:} For each pair $i$,
  Charlie performs a Bell measurement on $(A_2^{(i)}, B_2^{(i)})$
  and announces the outcome to Alice and Bob. They confirm they
  received identical reports and abort otherwise, per (A1)--(A2).
  \item \textbf{Test selection:} Alice draws, under (A3), a
  uniformly random subset $T \subset \{1,\ldots,N\}$ of size $k$ with a
  uniformly random balanced partition $T = T_Z \sqcup T_X$,
  $|T_Z| = |T_X| = k/2$, and, independently of $T$, a uniform
  permutation $\pi$ of the remaining set $D$. She sends these to Bob.
  \item \textbf{Verification:} For each $b \in \{Z, X\}$ and each
  $i \in T_b$, Alice applies the announced correction $P_{m^{(i)}}$ to
  $A_1^{(i)}$; then Alice and Bob measure $A_1^{(i)}$ and $B_1^{(i)}$ in
  basis $b$ and compare their outcomes. They compute the
  empirical anti-correlated fraction $\hat{a}_b$ over $T_b$ and accept
  if
  \begin{equation*}
    \hat{a}_Z \leq \tfrac{1}{2}\gamma_Z + \theta
    \quad\text{and}\quad
    \hat{a}_X \leq \tfrac{1}{2}\gamma_X + \theta,
  \end{equation*}
  where $\theta$ is fixed in~\cref{eq:budget} below; otherwise they
  abort ($\bot$).
  \item \textbf{Use:} On acceptance, Alice applies $P_{m^{(i)}}$ to
  $A_1^{(i)}$ for each of the $M$ pairs in $D$, and these are passed to
  the application in the order $\pi$
  (\Cref{rem:symmetrisation}).
\end{enumerate}

The correction is the only use Alice and Bob make of $m^{(i)}$, so it is
the only route by which a misreport can reach the test. 
An announced
$m^{(i)} \neq r^{(i)}$ leaves a residual $P_{e}$ on the pair. 
On a test pair no gate need be applied, since a Pauli commutes past
a $Z$ or $X$ measurement as a relabelling of Alice's outcome bit.

\begin{remark}\label{rem:isolation}
Incoming qubits must not be admitted to the application's quantum
memory until verification passes: if they share hardware with qubits
holding sensitive data, cross-talk can leak information to adjacent
qubits~\cite{sarovar2020detecting,murali2020software,tan2025know}.
Step~5 forwards pairs only after acceptance, in line with the
observation that \textit{``the qubits used for certification must be
isolated from the external network''}~\cite{satoh2021attacking}.
\end{remark}

The target $\ket{\phi^+}$ is correlated in both bases, so the test
looks at the probability $a_b$ of an \emph{anti-correlated} outcome in
basis $b$. On an honest execution $a_b = \gamma_b/2$, which defines the
baselines; they may differ (on dephasing-dominated hardware
$\gamma_X \geq \gamma_Z$), and we write
$\gamma_{\max} = \max\{\gamma_Z,\, \gamma_X\}$. Nothing is lost by
ignoring the rest of the joint distribution, since every deviation moves
mass only between correlated and anti-correlated outcomes. The baselines relate to the infidelity $\delta$
as follows.

\begin{lemma}\label{lem:delta-gamma}
Let $\rho_{AB}$ have fidelity $F(\rho_{AB},\ket{\phi^+}\bra{\phi^+})
= 1-\delta$, and let $\gamma_Z, \gamma_X$ be the baselines induced by
$\rho_{AB}$ in the sense above. Then
\begin{equation}\label{eq:delta-gamma}
  \delta \;\leq\; \frac{\gamma_Z + \gamma_X}{2} \;\leq\; 2\delta .
\end{equation}
\end{lemma}
\begin{proof}
Let $p_{\phi^+}, p_{\phi^-}, p_{\psi^+}, p_{\psi^-}$ denote the
diagonal weights of $\rho_{AB}$ in the Bell basis, so
$\delta = 1 - p_{\phi^+}$. Bell coherences do not affect the
quantities of interest: for the $Z$ basis, the $\ket{\phi^+}\bra{\phi^-}$
term contributes $+\tfrac{1}{2}$ to $\bra{00}\rho\ket{00}$ and
$-\tfrac{1}{2}$ to $\bra{11}\rho\ket{11}$, so it cancels in the total
correlated probability, and similarly in the $X$ basis. Hence the
anti-correlated probability, which is $\gamma_b/2$ by definition of
$\gamma_b$, is
\begin{equation}
  \tfrac{1}{2}\gamma_Z = p_{\psi^+} + p_{\psi^-},
  \qquad
  \tfrac{1}{2}\gamma_X = p_{\phi^-} + p_{\psi^-},
\end{equation}
since $\ket{\psi^\pm}$ are anti-correlated in $Z$ while
$\ket{\phi^-},\ket{\psi^-}$ are anti-correlated in $X$. Adding,
\begin{equation}
  \frac{\gamma_Z+\gamma_X}{2}
  = p_{\phi^-} + p_{\psi^+} + 2p_{\psi^-}
  = \delta + p_{\psi^-},
\end{equation}
and $0 \leq p_{\psi^-} \leq \delta$ gives~\cref{eq:delta-gamma}.
Equality on the left holds when $p_{\psi^-} = 0$ and on the right when
$p_{\psi^-} = \delta$, i.e.\ when the entire infidelity is carried by
$\ket{\psi^-}$, the Bell state visible in both bases.
\end{proof}

For the depolarising channel $(\gamma_Z+\gamma_X)/2 = 4\delta/3$. Both
bounds are attained, because fidelity does not fix how
the infidelity is spread over the Bell basis, and $\ket{\psi^-}$ is
counted by both bases.

By \Cref{lem:attack}, a deviation consists of Charlie announcing
$r \oplus e$ in place of $r$, which applies
$P_e \in \{\sigma_x, \sigma_z, \sigma_x\sigma_z\}$ to the delivered
pair. Let $f_x, f_z, f_y$ denote the fractions of pairs carrying each
of the three deviations, and let $f_N = f_x + f_z + f_y$ be the total
deviation rate over the batch of $N$ pairs, from which the
delivered-pair rate $f$ is recovered in~\cref{eq:transfer}. Each
deviation maps $\ket{\phi^+}$ to another Bell state, which in a given
basis is either correlated as the target is or anti-correlated.

A deviation is visible in basis $b$ if it flips the outcome
there; let $f_Z = f_x + f_y$ and $f_X = f_z + f_y$ be the fractions of
pairs visible in each basis. A visible pair is anti-correlated with
probability $1-\gamma_b/2$ instead of $\gamma_b/2$, so
\begin{equation}\label{eq:per-basis-a}
  a_b \;=\; (1-f_b)\tfrac{\gamma_b}{2}
    + f_b\bigl(1-\tfrac{\gamma_b}{2}\bigr)
    \;=\; \tfrac{\gamma_b}{2} + f_b\,(1-\gamma_b).
\end{equation}
The test detects the excess $f_b(1-\gamma_b)$. Since acceptance requires
both bases to pass, Charlie's best allocation minimises the larger of
the two excesses.

\begin{lemma}\label{lem:visibility}
For any allocation $(f_x, f_z, f_y)$ with $f_x + f_z + f_y = f_N$,
\begin{equation}\label{eq:worstcase}
  \max_{b \in \{Z,X\}} \Bigl(a_b - \tfrac{\gamma_b}{2}\Bigr)
  \;\geq\; \frac{f_N}{2}\,\bigl(1 - \gamma_{\max}\bigr),
\end{equation}
and the bound is attained, when $\gamma_Z = \gamma_X$, by the balanced
mixture $f_x = f_z = f_N/2$, $f_y = 0$.
\end{lemma}
\begin{proof}
The maximum of two quantities is at least their mean, so
by~\cref{eq:per-basis-a},
\begin{equation}
\begin{split}
  \max_b \Bigl(a_b - \tfrac{\gamma_b}{2}\Bigr)
  &\geq \tfrac{1}{2}\bigl[f_Z(1-\gamma_Z) + f_X(1-\gamma_X)\bigr] \\
  &\geq \tfrac{1}{2}(1-\gamma_{\max})\,(f_Z + f_X).
\end{split}
\end{equation}
Now $f_Z + f_X = (f_x + f_y) + (f_z + f_y) = f_N + f_y \geq f_N$,
which gives~\cref{eq:worstcase}. For attainment, take
$\gamma_Z = \gamma_X$ and $f_x = f_z = f_N/2$, $f_y = 0$: then
$f_Z = f_X = f_N/2$ and both sides equal
$\tfrac{f_N}{2}(1-\gamma_{\max})$.
\end{proof}

The factor $\tfrac{1}{2}$ in~\cref{eq:worstcase} comes from testing in
two bases: an adversary who splits his deviations evenly
between $\sigma_x$ and $\sigma_z$ is seen by only half of the test set
in either basis. That the worst case is a deviation spread thinly
rather than concentrated mirrors the finding of Wiesner et al.\ in a
different setting~\cite{wiesner2025cutandchoose}, where an i.i.d.\ attack
deviating slightly everywhere beats one that corrupts a single round.

\begin{remark}[Index-dependent honest noise]\label{rem:index-noise}
Pairs prepared early are held longer and decohere more
(\Cref{ssec:memory}). Let $\gamma^{(i)}_b$ denote the honest rate on pair
$i$, and $\gamma_b$ the batch average declared under (A6). The acceptance
test needs only the average, since \Cref{lem:conc} allows arbitrary
per-pair rates. But the factor $(1-\gamma_b)$ in~\cref{eq:per-basis-a}
converts deviation into signal on the pairs Charlie chose to corrupt,
\begin{equation}\label{eq:index-excess}
  a_b - \frac{\gamma_b}{2}
  \;=\; \frac{1}{N}\sum_{i \in V_b}\bigl(1-\gamma^{(i)}_b\bigr),
\end{equation}
with $V_b$ the corrupted set visible in basis $b$, and he will choose
the noisiest. \Cref{eq:worstcase,eq:phi-bound} are therefore to be read
with $\gamma_{\max} \geq \max_{i,b}\gamma^{(i)}_b$, and this per-pair
reading is the one used throughout \Cref{sec:resources}, where
\Cref{ssec:lifetime} makes the contribution of storage explicit. Nothing else in
the analysis needs the noise to be identical across pairs. 
The misreports that are easiest to hide also do the least damage: for a depolarising pair,
$1-\gamma^{(i)}_b$ is the fidelity a misreport destroys.
\end{remark}

\subsection{Security analysis}\label{sec:protocol-analysis}

\begin{figure*}[t]
\centering
\begin{tikzpicture}[
  >=Latex,
  every node/.style={font=\footnotesize},
  box/.style={draw, rounded corners=2pt, align=center, inner sep=5pt,
              fill=gray!5},
  goal/.style={draw, rounded corners=2pt, align=center, inner sep=5pt,
              fill=gray!12, very thick},
  note/.style={font=\scriptsize, align=left, inner sep=2pt},
]
\newcommand{\costs}[1]{{\color{gray!45!black}#1}}

\node[box] (obs) at (0,0)
  {measured on the $k$ test pairs (step 4):\quad
   $\hat{a}_Z,\ \hat{a}_X$};

\coordinate (j)  at (0,-1.5);
\draw[thick] (obs) -- (j);

\node[box] (live) at (-4.2,-3.5)
  {\textbf{liveness}\\[1pt]
   $\Pr(\bot \mid \text{honest}) \leq \tfrac{1}{3}\varepsilon_{\mathrm{sec}}$};
\draw[->,thick] (j) -| (live);
\node[note, anchor=north] at (-2.45,-1.62)
  {honest execution\\ concentrates at $\tfrac{\gamma_b}{2}$\\
   \costs{costs $\varepsilon_{\mathrm{sec}}/3$}};

\node[box]  (exc) at (3.0,-3.5)
  {$a_b - \tfrac{\gamma_b}{2} \;\leq\; \theta + \nu$};
\node[box]  (bat) at (3.0,-5.6)
  {batch rate\quad
   $f_N \;\leq\; \dfrac{2(\theta+\nu)}{1-\gamma_{\max}}$};
\node[goal] (del) at (3.0,-7.9)
  {delivered rate\quad $f \;\leq\; f_{\mathrm{target}}$\\[1pt]
   \textbf{\Cref{thm:security}}};

\draw[->,thick] (j) -| (exc);
\draw[->,thick] (exc) -- (bat);
\draw[->,thick] (bat) -- (del);

\node[note, anchor=north] at (0.95,-1.62)
  {if accepted: $\hat{a}_b \leq \tfrac{\gamma_b}{2}+\theta$\\
   \Cref{lem:conc} adds margin $\nu$\\
   \costs{costs $\varepsilon_{\mathrm{sec}}/3$}};
\node[note, anchor=west] at (3.3,-4.55)
  {\Cref{lem:visibility}: worst case\\ over $(f_x,f_z,f_y)$\\
   \costs{spends none}};
\node[note, anchor=west] at (3.3,-6.75)
  {\cref{eq:hypergeom}: cut-and-choose\\ transfer, tolerance $t$\\
   \costs{costs $\varepsilon_{\mathrm{sec}}/3$}};

\end{tikzpicture}
\caption{The route from the two measured fractions to
\Cref{thm:security}, with the share of $\varepsilon_{\mathrm{sec}}$ each
step uses. The left branch bounds liveness on an honest execution, the
right security on an accepting one.}
\label{fig:chain}
\end{figure*}
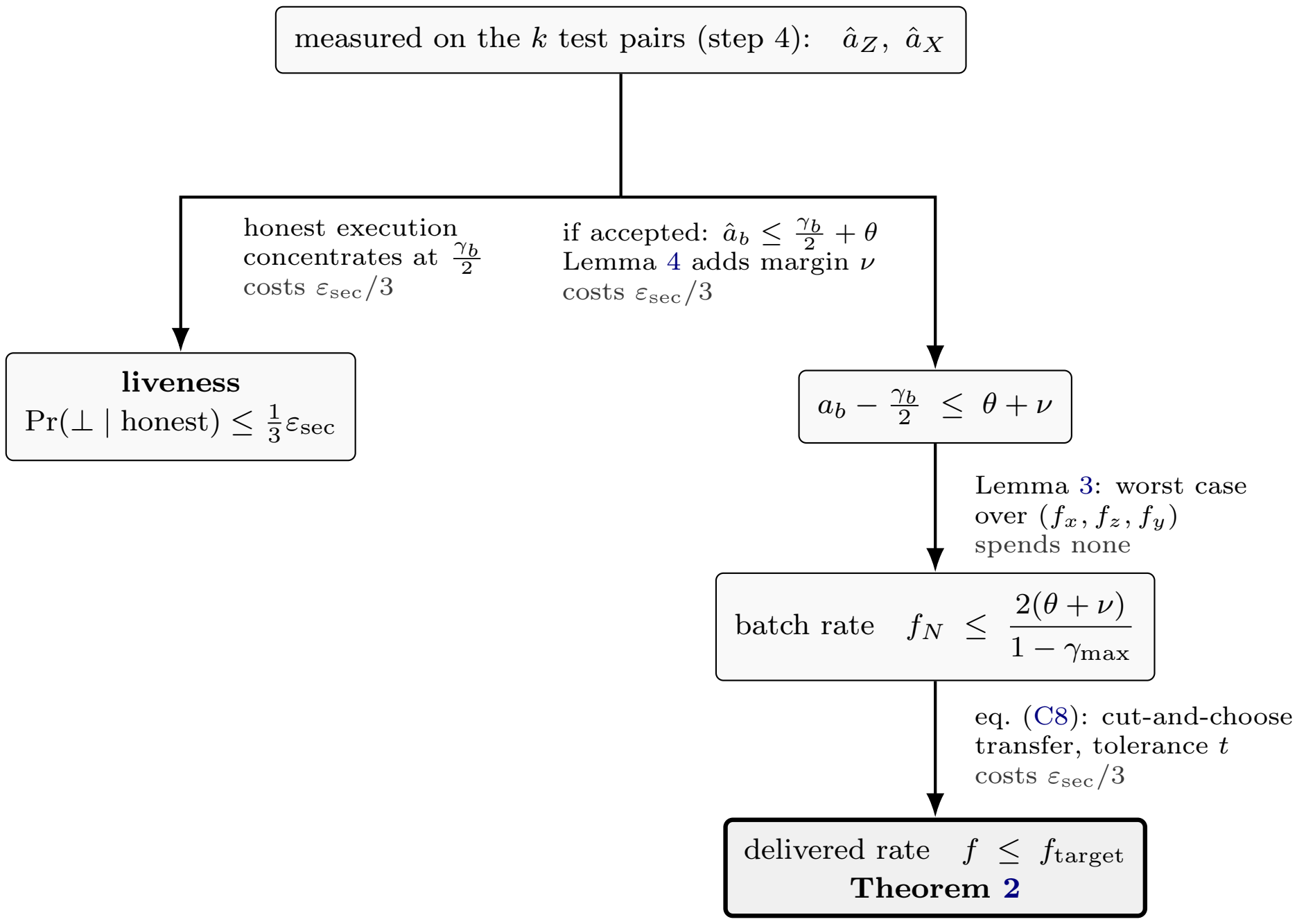

An execution can fail in two ways: \emph{false rejection} of an honest
execution, a liveness failure, or \emph{false acceptance} with
$f > f_{\mathrm{target}}$, a security failure. \Cref{fig:chain} shows
the route from the measured fractions to \Cref{thm:security}.

Both failure bounds rest on a concentration statement for any sample
drawn without replacement whose members then return noisy outcomes.

\begin{lemma}\label{lem:conc}
Let $p_1,\dots,p_N \in [0,1]$, let $T_b \subseteq \{1,\dots,N\}$ be a
uniformly random subset of size $n$, and, conditionally on $T_b$, let
$X_i$ for $i \in T_b$ be independent with
$X_i \sim \mathrm{Bernoulli}(p_i)$. Let
$\bar{p} = \frac{1}{N}\sum_{i=1}^{N}p_i$ and
$\hat{p} = \frac{1}{n}\sum_{i \in T_b}X_i$. Then for every
$\lambda \in \mathbb{R}$,
\begin{equation}\label{eq:conc-mgf}
  \mathbb{E}\bigl[e^{\lambda n \hat{p}}\bigr]
  \;\leq\; \bigl(1 - \bar{p} + \bar{p}\,e^{\lambda}\bigr)^{n},
\end{equation}
the moment generating function of a $\mathrm{Binomial}(n,\bar{p})$
variable. Consequently every Chernoff bound available for $n$
independent $\mathrm{Bernoulli}(\bar{p})$ draws holds for $\hat{p}$
unchanged, in both tails; in particular, for any $t>0$,
\begin{equation}\label{eq:conc}
  \Pr\big(\hat{p} - \bar{p} > t\big)
  \;\leq\; \exp\!\left(-\frac{n t^2}{2\bar{p}(1-\bar{p}) + \tfrac{2}{3}t}\right),
\end{equation}
and symmetrically in the lower tail.
\end{lemma}
\begin{proof}
Fix $\lambda \in \mathbb{R}$ and put
$g_i = 1 - p_i + p_ie^{\lambda} > 0$, the moment generating function of
$X_i$ at $\lambda$. Conditioning on $T_b$ and using the independence of
the $X_i$ given it,
\begin{equation}
  \mathbb{E}\bigl[e^{\lambda n\hat{p}}\bigr]
  \;=\; \mathbb{E}_{T_b}\Bigl[\,\prod_{i \in T_b} g_i\Bigr]
  \;=\; \binom{N}{n}^{-1} e_n(g_1,\dots,g_N),
\end{equation}
since averaging $\prod_{i \in T_b}g_i$ over all $n$-subsets is by
definition the normalised elementary symmetric polynomial of degree
$n$. This is Hoeffding's comparison between sampling without and with
replacement~\cite[Thm~4]{hoeffding1963probability}, applied to the
population $\{\log g_i\}_{i=1}^N$ and the convex function $\exp$;
equivalently, Maclaurin's inequality~\cite{hardy1952inequalities} for
positive reals gives
\begin{equation}
  \left(\binom{N}{n}^{-1}e_n(g_1,\dots,g_N)\right)^{1/n}
  \;\leq\; \frac{1}{N}\sum_{i=1}^{N}g_i
  \;=\; 1 - \bar{p} + \bar{p}\,e^{\lambda},
\end{equation}
which is~\cref{eq:conc-mgf}. Since the bound holds at every $\lambda$,
of either sign, the Chernoff argument for a
$\mathrm{Binomial}(n,\bar{p})$ variable goes through verbatim with
$\hat{p}$ in place of the binomial mean, in both tails.
\Cref{eq:conc} is the Bernstein form of that argument, obtained by
bounding the moment generating function using the variance
$\bar{p}(1-\bar{p})$ rather than its maximum
$1/4$~\cite{hoeffding1963probability,chernoff1952measure}.
\end{proof}

\Cref{lem:conc} follows directly from Hoeffding's comparison theorem.
We state it separately for two reasons. First, it handles our situation
in a single step: a sample drawn without replacement, whose members then
produce outcomes at rates of their own. The usual approach bounds the
sampling and the measurement separately and combines them with a union
bound. Second, it bounds the moment generating function itself rather
than a single tail probability, so any of the standard exponential
bounds on large deviations can be derived from it, including the
variance-aware one we use.

Using the variance, rather than its worst case, makes a large
difference here. The rate being estimated is close to $\gamma_b/2$, at
most a few per cent, so its variance is far below the worst-case value
of $1/4$. This reduces the number of test pairs needed by more than an
order of magnitude (\Cref{ssec:test-overhead}).

We control three events, and assign each a third of the failure budget:
\begin{enumerate}
  \item an honest execution produces an $\hat{a}_b$ more than $\theta$
  above $\gamma_b/2$ in either basis (\Cref{sssec:false-rejection});
  \item some $\hat{a}_b$ underestimates the true $a_b$ by more than
  $\nu$;
  \item the test set $T$ happens to under-represent the deviated pairs
  by more than $t$ (\Cref{sssec:false-acceptance}).
\end{enumerate}
With $L = \log(6/\varepsilon_{\mathrm{sec}})$ (natural logarithm), $n = k/2$ the number of pairs
tested in each basis, and $g = \gamma_{\max}/2$ the honest
anti-correlation rate in the noisier basis, the acceptance threshold
$\theta$, the estimation precision $\nu$ and the sampling
tolerance $t$ take the values
\begin{align}\label{eq:budget}
  \theta &\;=\; \frac{L}{3n}
    + \sqrt{\frac{L^2}{9n^2} + \frac{2g(1-g)L}{n}}, \notag\\
  \nu &\;=\; \frac{4L}{3n}
    + \sqrt{\frac{16L^2}{9n^2} + \frac{2(g+\theta)L}{n}}, \\
  t &\;=\; \sqrt{\Bigl(1-\frac{k-1}{N}\Bigr)\frac{L}{2k}} . \notag
\end{align}
The first two are instances of
\Cref{lem:conc} with $n = k/2$; the values above are the positive roots
of the quadratics that setting~\cref{eq:conc} equal to $\varepsilon_{\mathrm{sec}}/6$
produces, so each event costs $\varepsilon_{\mathrm{sec}}/6$ per basis and $\varepsilon_{\mathrm{sec}}/3$ over
the two. The third is bounded in~\cref{eq:hypergeom}
(Appendix~\ref{app:security-analysis}). A union
bound gives a total failure probability of at most $\varepsilon_{\mathrm{sec}}$.
Each margin has two parts, and which one dominates depends on $ng$, the
number of anti-correlated outcomes an honest relay is expected to produce
among the $n$ pairs tested in a basis. When $ng \gg L$, the honest count
is close to Gaussian, and both margins are close to
$\sqrt{2L}\cdot\sqrt{g/n}$: a fixed number $\sqrt{2L}$ of standard
deviations of the honest fraction, about $3.6$ at
$\varepsilon_{\mathrm{sec}} = 10^{-2}$. When $ng \lesssim L$, too few
honest anti-correlations are expected for that approximation, and the
margins are set instead by the $L/n$ terms; as $g \to 0$,
$\theta = 2L/3n$ and $\nu = (4+2\sqrt{7})L/3n \approx 3.1\,L/n$. These
two regimes give the two scalings of $k$ in~\cref{eq:k-regimes}. At the
operating point of
\Cref{ssec:test-overhead}, $ng \approx 2{,}100$ against $L \approx 6.4$,
so the first regime applies and $\theta \approx \nu \approx 2\times10^{-3}$.
In both regimes $\nu$ is the larger, because the true rate it must cover
is unknown: bounding $a_b(1-a_b) \leq a_b \leq g + \theta + \nu$ to close
the argument is what turns the $2/3$ in $\theta$ into the $8/3$
behind $\nu$.

\subsubsection{Outline of the derivation}\label{sssec:outline}

We outline the two failure bounds along \Cref{fig:chain}; the
derivations are in Appendix~\ref{app:security-analysis}.

\emph{False rejection.} In an honest execution the mean anti-correlation
rate in basis $b$ is $\gamma_b/2$, so \Cref{lem:conc} bounds the
probability that $\hat{a}_b$ exceeds $\gamma_b/2 + \theta$ by
$\varepsilon_{\mathrm{sec}}/6$ in each basis. An honest Charlie is
therefore rejected with probability at most
$\varepsilon_{\mathrm{sec}}/3$ (\cref{eq:false-reject}).

\emph{False acceptance.} Outside the second event, acceptance implies
that the true rate exceeds the baseline by at most $\theta + \nu$ in both
bases, and \Cref{lem:visibility} turns this into a bound on the batch
rate, $f_N \leq 2(\theta+\nu)/(1-\gamma_{\max})$. Outside the third, the
test set represents the deviated pairs fairly, so the bound carries over
to the delivered pairs at an additive cost of $(k/M)\,t$:
\begin{equation}\label{eq:f-final}
  f \;\leq\; \frac{2}{1-\gamma_{\max}}
  \Bigl[\, \theta + \nu + \frac{k}{M}\,t \,\Bigr].
\end{equation}
Requiring the right-hand side to be at most $f_{\mathrm{target}}$ gives
the condition~\cref{eq:n-threshold} on $k$.

\subsubsection{Total failure probability}

The two failure modes concern different executions, so their
probabilities are not added. False rejection concerns an honest
execution: by~\cref{eq:false-reject}, it is rejected with probability
at most $\varepsilon_{\mathrm{sec}}/3$. False acceptance concerns a
dishonest one. Outside the second and third events listed
before~\cref{eq:budget}, acceptance implies
$f \leq f_{\mathrm{target}}$ (\Cref{sssec:false-acceptance}), so
\begin{equation}
  \Pr[\mathsf{acc} \wedge f > f_{\mathrm{target}}]
  \;\leq\; \tfrac{2}{3}\,\varepsilon_{\mathrm{sec}} .
\end{equation}

\begin{remark}\label{rem:two-budgets}
Since a rejected honest execution only costs a retry, false rejection
could be given its own parameter $\varepsilon_{\mathrm{live}}$. This
replaces $L$ by $\log(4/\varepsilon_{\mathrm{sec}})$ in $\nu$ and $t$
and by $\log(2/\varepsilon_{\mathrm{live}})$ in $\theta$, saving a few
test pairs; we use a single parameter for simplicity.
\end{remark}

\subsection{The security guarantee}\label{ssec:guarantee}

\begin{theorem}\label{thm:security}
Given pre-agreed parameters $f_{\mathrm{target}}$, $\gamma_Z$,
$\gamma_X$, $\varepsilon_{\mathrm{sec}}$ and $M$, the two-basis protocol is
$(f_{\mathrm{target}}, \varepsilon_{\mathrm{sec}})$-secure for any even $k$ satisfying
\begin{equation}\label{eq:k}
\begin{split}
  &2\bigl(\theta + \nu\bigr)
  + \frac{2k}{M}\sqrt{\Bigl(1-\frac{k-1}{N}\Bigr)
              \frac{\log(6/\varepsilon_{\mathrm{sec}})}{2k}} \\
  &\qquad\leq\; f_{\mathrm{target}}\,(1-\gamma_{\max}),
\end{split}
\end{equation}
with $\theta$ and $\nu$ as in~\cref{eq:budget}, $N = M+k$, and $f$
the deviation rate on the $M$ delivered pairs in the sense of
\Cref{def:prob-security}. Provided the declared baselines are correct,
it is complete: an honest Charlie is accepted except with probability
$\varepsilon_{\mathrm{sec}}/3$.
\end{theorem}
\begin{proof}
Completeness is~\cref{eq:false-reject} at the budgeted $\theta$, and
holds at $\Delta_b = 0$ in the sense of \Cref{lem:misspec}. Soundness is
the chain of \Cref{sssec:false-acceptance}, whose
conclusion~\cref{eq:n-threshold} is~\cref{eq:k}; the two events it
conditions on (items~2 and~3 above) fail with total probability at most
$\tfrac{2}{3}\varepsilon_{\mathrm{sec}}$.
\end{proof}

\Cref{thm:security} counts misreports, but the test itself only sees
anti-correlations above the declared baselines. Any extra Pauli error on
the pairs, applied independently of the test set, is counted the same
way, whether it comes from a lie, a degraded measurement or a drifting
memory. It is the threat model that lets us call this excess a
misreport rate, and \Cref{prop:unidentifiable} shows that no test can
separate the two.

The left-hand side of~\cref{eq:k} is not monotone in $k$: the first term
falls and the sampling transfer rises, so the feasible $k$ form an
interval, and the smallest is found on the falling branch.

\Cref{def:prob-security} bounds a rate of misreports, a classical
quantity, whereas an application uses the delivered quantum state. The
following corollary bounds that state.

\begin{corollary}\label{cor:state}
Let $\tilde{\rho}_D$ denote the state
delivered on the $M$ retained pairs restricted to the accepting branch,
that is the sub-normalised operator with
$\operatorname{tr}\tilde{\rho}_D = \Pr[\mathsf{acc}]$, and
write $\mathcal{P}_w$ for the set of Pauli operators on Alice's $M$
halves acting non-trivially on at most $w = f_{\mathrm{target}}M$ of
them. Then there are weights $q(E) \geq 0$ on $\mathcal{P}_w$ with
$\sum_{E} q(E) \leq \Pr[\mathsf{acc}]$ and
\begin{equation}\label{eq:cor-state}
  \Bigl\|\; \tilde{\rho}_D \;-\; \sum_{E \in \mathcal{P}_w} q(E)\,
  (E \otimes I)\,\rho^{\mathrm{hon}}_D\,(E\otimes I)^\dagger
  \;\Bigr\|_1 \;\leq\; \tfrac{2}{3}\varepsilon_{\mathrm{sec}} ,
\end{equation}
where $\rho^{\mathrm{hon}}_D$ is the state that would have been output if
Charlie reported truthfully.
\end{corollary}
\begin{proof}
Condition on Charlie's randomness and on the test set, so that his
deviation pattern is a fixed labelling $e_1,\dots,e_M$ of the delivered
pairs. By property~3 of \Cref{lem:attack}, a misreport on pair $i$
delivers $(P_{e_i}\otimes I)\rho^{\mathrm{hon}}_i(P_{e_i}\otimes
I)^\dagger$ and a truthful report delivers $\rho^{\mathrm{hon}}_i$, so
on that branch the delivered state is
$(E\otimes I)\rho^{\mathrm{hon}}_D(E\otimes I)^\dagger$ with
$E = \bigotimes_{i \in D}P_{e_i}$ and $e_i = 00$ where the report was
truthful. The weight of $E$ is the number of misreported delivered
pairs, namely $fM$. Let $p_E$ denote the probability that the execution
accepts \emph{and} leaves pattern $E$; summing over the conditioning,
\begin{equation}\label{eq:cor-mixture}
  \tilde{\rho}_D = \sum_{E} p_E\,
  (E\otimes I)\,\rho^{\mathrm{hon}}_D\,(E\otimes I)^\dagger ,
  \quad \sum_{E} p_E = \Pr[\mathsf{acc}] .
\end{equation}
Restricting to the accepting branch does not disturb
$\rho^{\mathrm{hon}}_D$ itself: acceptance is determined by the test
pairs and by the deviation pattern, and by (A7) the noise on the
delivered pairs is independent of the noise on the test pairs.

Take $q$ to be the restriction of $p$ to $\mathcal{P}_w$, so that
$\sum_E q(E) \leq \sum_E p_E = \Pr[\mathsf{acc}]$. The discarded terms
of~\cref{eq:cor-mixture} are positive semidefinite with traces $p_E$,
so their sum has trace norm $\sum_{E \notin \mathcal{P}_w} p_E$, which
is the probability that the protocol accepts with
$f > f_{\mathrm{target}}$. By \Cref{thm:security} that is at most
$\tfrac{2}{3}\varepsilon_{\mathrm{sec}}$, which
is~\cref{eq:cor-state}.
\end{proof}

Since Charlie's pattern also decides acceptance, only the mixture
in~\cref{eq:cor-state}, not a single fixed Pauli, relates the delivered
state to the honest one, and any subsequent error correction must handle 
this mixture. The state conditioned on acceptance
obeys the same bound with $\tfrac{2}{3}\varepsilon_{\mathrm{sec}}$
replaced by $\tfrac{4}{3}\varepsilon_{\mathrm{sec}}/\Pr[\mathsf{acc}]$,
which is immaterial against an honest Charlie.

\begin{remark}[Symmetrisation]\label{rem:symmetrisation}
\Cref{cor:state} bounds the weight of $E$ but not its support: Charlie
fixes his pattern before $T$ is drawn, so all $f_{\mathrm{target}}M$
corrupted pairs could sit at the start of the delivered stream.
Delivering in the order $\pi$ (step~5) is the symmetrisation step
standard in QKD security
proofs~\cite{shor2000simple,renner2008security,christandl2009postselection}
and in distillation through untrusted
repeaters~\cite{aschauer2002private}. By (A3),
$\pi$ is drawn after Charlie commits and independently of $T$, while
acceptance depends only on $T$ and his pattern; so conditioned on
acceptance and on the pattern, $\pi$ is still uniform, and the support of
$E$ is a uniformly random $fM$-subset of $D$. The permutation leaves
\Cref{thm:security} unchanged, but places the corrupted pairs in
uniformly random positions, as an error-correcting decoder and the
distillation analysis of \Cref{ssec:distil-compose} assume.

The permutation need not be secret. Charlie reads $\pi$ from the
authenticated but public channel of (A3), and so learns which delivered
pairs carry which Pauli; but the positions are uniformly random whether
or not he knows them, and the delivered state is the same either way.
The knowledge helps only an adversary who can act again on the delivered
pairs, which lies outside our threat model, whereas keeping $\pi$ secret
would cost $\log_2 M! \approx 1.8\times10^{7}$ bits of pre-shared key at
$M = 10^6$. For the same reason the ideal functionality of
\Cref{def:F-relaxed} gives $\pi$ to the adversary, so that the simulator
of \Cref{thm:composable} can reproduce what a real Charlie reads from
the channel.
\end{remark}

\subsection{A fidelity bound for an arbitrary relay}\label{ssec:fidelity-floor}

\Cref{thm:security} and \Cref{cor:state} rely on the threat model of
\Cref{ssec:threat-scope}. The same test also gives a weaker guarantee
that holds without it.

\begin{proposition}\label{prop:fidelity}
Let $\rho$ be the corrected state of a delivered pair, however it arose,
write $F = \bra{\phi^+}\rho\ket{\phi^+}$, and let $a_Z, a_X$ be the
probabilities that a $Z$- or $X$-basis measurement of it returns an
anti-correlated outcome. Then
\begin{equation}\label{eq:fid-bound}
  a_Z + a_X \;=\; 1 - F + \bra{\psi^-}\rho\ket{\psi^-}
  \;\geq\; 1 - F .
\end{equation}
Consequently, if Charlie applies an arbitrary instrument to each
$(A_2^{(i)}, B_2^{(i)})$, independently across pairs and fixed before
$T$ is drawn, acceptance implies that the batch-average fidelity
$\bar{F}$ satisfies
\begin{equation}\label{eq:fid-accept}
  1 - \bar{F} \;\leq\; \tfrac{1}{2}(\gamma_Z + \gamma_X)
  + 2(\theta + \nu)
\end{equation}
except with probability $\tfrac{2}{3}\varepsilon_{\mathrm{sec}}$.
\end{proposition}
\begin{proof}
The identity is the computation in \Cref{lem:delta-gamma}, which assumes
nothing about how $\rho$ arose: Bell coherences contribute to neither
diagonal, so $a_Z = p_{\psi^+} + p_{\psi^-}$ and
$a_X = p_{\phi^-} + p_{\psi^-}$, and the four weights sum to one. For
the second claim, \Cref{lem:conc} bounds each basis average
$\bar{a}_b \leq \hat{a}_b + \nu$ whatever the per-pair rates, and
acceptance bounds $\hat{a}_b$ by $\gamma_b/2 + \theta$. Both
$\bar{a}_Z$ and $\bar{a}_X$ are averages over the same $N$ pairs, so
summing them and applying~\cref{eq:fid-bound} to each pair gives
\cref{eq:fid-accept}. Only the two estimation events are used, so
$\varepsilon_{\mathrm{sec}}/3$ would suffice; we quote
$\tfrac{2}{3}\varepsilon_{\mathrm{sec}}$ to match \Cref{thm:security}.
\end{proof}

Acceptance thus bounds the fidelity of the delivered pairs for any
per-pair strategy of Charlie, but not the misreport rate; that stronger
statement (\Cref{thm:security}) needs the threat model. A Charlie who measures
$A_2$ and $B_2$ separately in the $Z$ basis leaves a product state with
$F = 1/2$, and~\cref{eq:fid-bound} is tight.

\subsection{Scaling of the test cost}\label{ssec:asymptotics}

Up to constants, the number of test pairs grows as
\begin{equation}\label{eq:k-scaling}
  k \;=\; O\!\left(\frac{(\gamma_{\max} + f_{\mathrm{target}})
              \log(1/\varepsilon_{\mathrm{sec}})}{f_{\mathrm{target}}^2}\right).
\end{equation}
On a clean channel this is $O(L/f_{\mathrm{target}})$; once the minimum noise
level dominates, it is $O(\gamma_{\max}L/f_{\mathrm{target}}^2)$. A
bound that ignores the variance, such as Hoeffding's, gives the second,
more expensive scaling whatever $\gamma_{\max}$ is. This is where the
variance-aware margins of \Cref{lem:conc} save test pairs.

Two approximations make this precise. Neglecting the sampling transfer
term in~\cref{eq:k} and writing
$\beta = f_{\mathrm{target}}(1-\gamma_{\max})$,
\begin{equation}\label{eq:k-regimes}
\begin{aligned}
  k \;&\approx\; \frac{32\,\gamma_{\max}L}{\beta^2}
    &&\text{when } k \gg L/\gamma_{\max}, \\[2pt]
  k \;&\approx\; \frac{4\bigl(6+2\sqrt{7}\bigr)L}{3\beta}
    \;\approx\; \frac{15.1\,L}{\beta}
    &&\text{as } \gamma_{\max} \to 0 .
\end{aligned}
\end{equation}
The second is exact in its limit, where $\theta = 2L/3n$ and
$\nu = (4+2\sqrt{7})L/3n$, and reproduces the $\gamma_{\max}=0$ column of
\Cref{tab:k-vs-f} at $f_{\mathrm{target}} = 0.3$, $0.1$ and $0.05$ to
within half a per cent. Neither is accurate at the operating point: the
first gives $1.1\times10^5$ against a tabulated $1.7\times10^5$, because
the neglected transfer term takes $14\%$ of the budget there.
\Cref{tab:k-vs-f} therefore reports exact solutions of~\cref{eq:k}.

\subsection{Sensitivity to calibration}\label{ssec:calibration}

The test separates Charlie's misreports from channel noise using the
declared baselines $\gamma_Z, \gamma_X$. If a declared baseline is
wrong, that separation shifts in one of two directions:
\begin{itemize}
\item \emph{Declared too low:} honest noise looks like misreporting, so
honest batches are rejected more often. Security is unaffected; only
liveness suffers.
\item \emph{Declared too high:} part of Charlie's misreporting hides
under the inflated minimum noise level and goes undetected. Security is lost.
\end{itemize}
The following lemma quantifies both.

\begin{lemma}\label{lem:misspec}
Suppose the analysis is run with assumed baselines $\gamma^a_b$ while
the true ones are $\gamma^t_b$, and let
$\Delta_b = \gamma^a_b - \gamma^t_b$. Then the observed excess over the
assumed baseline is
\begin{equation}\label{eq:misspec}
  a_b - \tfrac{\gamma^a_b}{2}
  \;=\; f_b\,(1-\gamma^t_b) \;-\; \tfrac{1}{2}\Delta_b .
\end{equation}
\end{lemma}
\begin{proof}
By~\cref{eq:per-basis-a} with the true baseline in place,
$a_b = \gamma^t_b/2 + f_b(1-\gamma^t_b)$. Subtract $\gamma^a_b/2$.
\end{proof}

\emph{Declared too low} ($\Delta_b < 0$). The offset
in~\cref{eq:misspec} now adds to the misreporting signal rather than
masking it. Acceptance implies
$f_b(1-\gamma^t_b) \leq \theta + \nu - \tfrac{1}{2}|\Delta_b|$, so the
guarantee of \Cref{thm:security} still holds. The cost is liveness: an
honest execution sits $\tfrac{1}{2}|\Delta_b|$ above the assumed
baseline, leaving only $\theta - \tfrac{1}{2}|\Delta_b|$ of room for
statistical fluctuation. Liveness therefore requires
\begin{equation}\label{eq:calib}
  |\Delta_b| \;\leq\; 2\theta ,
\end{equation}
but this is not enough. At $|\Delta_b| = 2\theta$ honest batches are
rejected about half the time, and any non-zero offset already raises
the false-rejection rate. The window is narrow because $\theta$ scales
as $\sqrt{\gamma_{\max}L/k}$, which is what keeps $k$ small: at the
operating point $2\theta \approx 0.4\%$, and honest batches are
rejected outright at $|\Delta_b| = 1\%$. A
deployment that cannot calibrate well within $\theta$ should set the
false-rejection probability separately (\Cref{rem:two-budgets}).

\begin{remark}[Declared too high]\label{rem:overest}
If $\Delta_b > 0$, the offset masks misreporting, and acceptance implies
only $f_b(1-\gamma^t_b) \leq \theta + \nu + \tfrac{1}{2}\Delta_b$.
Propagating this through \Cref{lem:visibility} as in~\cref{eq:f-final}
replaces the guarantee of \Cref{thm:security} by
\begin{equation}\label{eq:inflated}
  f \;\leq\; f_{\mathrm{target}}
   \;+\; \frac{\max_b \Delta_b}{1-\gamma_{\max}} .
\end{equation}
A minimum noise level declared $\Delta$ above the truth gives Charlie
$\Delta/(1-\gamma_{\max})$ of undetected misreporting. A provider
certifying $\gamma_Z, \gamma_X$ who also operates the relay, or is
merely lax, can thus widen his margin, and nothing in the protocol
detects it. Calibrating more carefully does not fix this, as the next
proposition shows.
\end{remark}

\begin{proposition}[Misreports are indistinguishable from readout
noise]\label{prop:unidentifiable}
Let $p = (p_e)_{e \in \{0,1\}^2}$ be a probability distribution, and
write $\Pi_r$ for the projector onto the Bell state labelled $r$.
Compare two relays that receive the same two qubits. The first is
honest, but its Bell measurement has readout noise: it implements the
POVM $\{E_m\}$ with
\begin{equation}
  E_m \;=\; \sum_{e} p_e\, \Pi_{m \oplus e},
\end{equation}
and announces its outcome $m$. The second performs the ideal Bell
measurement, obtains $r$, and announces $m = r \oplus e$ with $e$ drawn
from $p$ independently of everything else. Then for every joint state
the users prepare, the joint state of the announcement and of all
registers the users retain is the same for the two relays.
\end{proposition}
\begin{proof}
Let $\omega_{RQ}$ be the joint state of the registers $R$ the users
retain and the two qubits $Q$ they send. A measurement of $Q$ with POVM
$\{E_m\}$, after which $Q$ is discarded, leaves the classical-quantum
state $\sum_m \ket{m}\bra{m} \otimes
\operatorname{tr}_Q[(I_R \otimes E_m)\,\omega_{RQ}]$, whichever
instrument implements it: since $Q$ is consumed, only the POVM elements
matter. For the second relay the announced $m$ arises from every pair
$(r, e)$ with $r \oplus e = m$, so its state is
$\sum_m \ket{m}\bra{m} \otimes \sum_e p_e
\operatorname{tr}_Q[(I_R \otimes \Pi_{m\oplus e})\,\omega_{RQ}]$, which
is the same operator.
\end{proof}

The proposition holds for any states Alice and Bob send, so probe
states, decoys and the protocol's own test pairs cannot separate the two
relays either. To the users, a relay that lies on a random fraction of
rounds is the same as an honest relay whose readout fails at that rate.
Whether an error counts as a lie or as noise is decided by the baseline,
not by the data.

In practice, this means the part of $\gamma_Z, \gamma_X$ coming from
Charlie's measurement, and from the links into it, cannot be estimated
from protocol data. If it is calibrated by sending pairs through the
relay, Charlie can inflate it by misreporting during calibration, and
by~\cref{eq:inflated} this gives him $\Delta/(1-\gamma_{\max})$ of
undetected deviation later. This part of the minimum noise level has to come from
outside the data, either from an inspection of the relay's hardware by
someone with access to it, or as a figure the provider commits to. The
users can, and should, characterise the rest themselves: their own
preparation, storage and measurement.

\Cref{prop:fidelity} does not need a trusted baseline. Its proof
uses the declared baselines only in the acceptance threshold, so on
acceptance the batch-average infidelity is at most the declared minimum noise level
plus $2(\theta + \nu)$, whatever was declared and whatever caused the
errors. An application actually sees only this total.
\Cref{thm:security} goes one step further and says how much of it is
due to lies, and that step is only as good as the baseline.

\section{Composable security and distillation}\label{sec:composable}

\Cref{thm:security} bounds what a single execution certifies. This
section shows that its output can also be used inside larger protocols
when the honest hardware is noisy.

\subsection{A realisable relaxed functionality}\label{ssec:composable-relaxed}

The obstruction of \Cref{rem:simulator} is that the ideal-world
adversary has no influence over the delivered state while the real-world
one does. We give the ideal-world adversary the same influence, limited
to the misreport rate bounded by \Cref{thm:security}.

\begin{definition}[Relaxed ES functionality]\label{def:F-relaxed}
Let $\Lambda = (\Lambda_1,\dots,\Lambda_N)$ be channels on a two-qubit
register, $f \in [0,1]$, and $N = M+k$. The functionality
$\mathcal{F}^{M}_{\Lambda,f}$ has an interface with Alice, one with
Bob and one with the adversary, and carries in addition the test size
$k$ and the acceptance rule of step~4 of \Cref{sec:protocol}, namely
the baselines $\gamma_Z, \gamma_X$ and the margin $\theta$. Alice and
Bob submit $A_2^{(1)},\dots,A_2^{(N)}$ and $B_2^{(1)},\dots,B_2^{(N)}$,
retaining their halves; the submitted qubits are consumed, and as in
\Cref{def:ideal-F} their contents play no part in what is delivered.
Then
\begin{enumerate}
 \item $\mathcal{F}$ draws $r^{(1)},\dots,r^{(N)}$ uniformly from
 $\{0,1\}^2$ and sends them to the adversary;
 \item the adversary returns either $\bot$ or a report string
 $m^{(1)},\dots,m^{(N)}$; $\mathcal{F}$ sets
 $e_i = m^{(i)} \oplus r^{(i)}$;
 \item $\mathcal{F}$ broadcasts $m^{(1)},\dots,m^{(N)}$ to Alice and Bob;
 \item $\mathcal{F}$ draws, from randomness of its own, a uniform
 $T \subset \{1,\dots,N\}$ with $|T| = k$, a uniform balanced partition
 $T = T_Z \sqcup T_X$, and a uniform permutation $\pi$ of
 $D = \{1,\dots,N\} \setminus T$. It sends $T$, its partition and $\pi$
 to the adversary and outputs them to Alice and Bob;
 \item $\mathcal{F}$ consumes the halves Alice and Bob retained on the
 indices in $T$. In their place it draws the test outcomes from the law
 they would have had: for each $b \in \{Z,X\}$ and each $i \in T_b$,
 the distribution of a basis-$b$ measurement of
 $(P_{e_i}\otimes I)\,\Lambda_i(\ket{\phi^+}\!\bra{\phi^+})\,
 (P_{e_i}\otimes I)^\dagger$, which it knows in closed form because
 $\Lambda$ is a parameter and $e$ was fixed at step~2. It sends those
 outcomes to the adversary, forms the anti-correlated fractions
 $\hat{a}_Z, \hat{a}_X$, and outputs those to Alice and Bob;
 \item if the adversary returned $\bot$, or if
 $\hat{a}_Z > \tfrac{1}{2}\gamma_Z + \theta$ or
 $\hat{a}_X > \tfrac{1}{2}\gamma_X + \theta$, then $\mathcal{F}$ outputs
 $\bot$ at all three interfaces. Otherwise, with
 $E = \bigotimes_{i \in D} P_{e_i}$: if $E$ acts non-trivially on more
 than $fM$ of the pairs in $D$, $\mathcal{F}$ outputs $\bot$; if not, it
 writes
 $(E\otimes I)\bigl[\bigotimes_{i \in D}\Lambda_i
 (\ket{\phi^+}\!\bra{\phi^+})\bigr](E\otimes I)^\dagger$
 into the registers Alice and Bob retained on the indices in $D$, in the
 order given by $\pi$.
\end{enumerate}
\end{definition}

The honest interface is therefore the reports $m$, the test set $T$
with its partition, the permutation $\pi$, the fractions $\hat{a}_Z, \hat{a}_X$, the
accept-or-$\bot$ decision, and on acceptance the $M$ delivered registers
in the order $\pi$.

\begin{remark}
\label{rem:not-transcription}
$\mathcal{F}$ concedes the outcomes $r^{(i)}$, without which a report
is not a lie about anything; the choice of $m$, the object under study;
and the test outcomes, which a real Charlie reading the authenticated
channel of (A2) already holds once his report is fixed. Everything else
is $\mathcal{F}$'s: $e_i$ is computed from $m$ and $r$, as
\Cref{lem:attack} shows it is in a real execution, so a truthful pair
cannot be corrupted; $T$ and $\pi$ are drawn after the report, so
neither the delivered set nor the position a corrupted pair takes in it
is his to name, although he learns both, as a real Charlie reading the
public channel of (A3) does; and $\Lambda$ is fixed, so the minimum noise level
is not up to Charlie to declare. $\mathcal{F}$ holds no quantum register, sampling the
test statistics from a law known in closed form. The acceptance rule is
a parameter, like $\Lambda$ and $f$, since an ideal functionality must specify
how much noise it will tolerate.
\end{remark}

At $\Lambda=\mathrm{id}$ and $f=0$, \Cref{def:F-relaxed} reduces to
$\mathcal{F}_0$, up to its $N$-pair interface and the information it
hands the adversary: the weight condition forces $E = I$, and lies on tested
pairs never reach the delivered state. So without noise,
\Cref{thm:nogo} rules out $f = 0$. With noise, an
honest execution already differs from $\mathcal{F}_0$
(\Cref{rem:batched-reading}). 

\begin{proposition}[No realisation at $f = 0$ under
noise]\label{prop:f0-noisy}
Let every $\Lambda_i$ equal a fixed $\Lambda$ with baselines
$\gamma_Z, \gamma_X$, write $g_b = \gamma_b/2$, and let $\Pi$ be the
protocol of \Cref{sec:protocol} with parameters $M$, $k = 2n$ and
margin $\theta$. Write
\begin{equation}
  B_m(x;\, v) \;=\; \exp\!\left(-\frac{m x^2}{2v + \tfrac{2}{3}x}\right)
\end{equation}
for the Bernstein bound of~\cref{eq:conc}. Take any $s \in (0,\theta)$
such that $\phi = s/(1-\gamma_Z)$ makes $\phi N$ an integer, any
$t' > 0$, and any $\tau > 0$ with $\tau < s_D$, where
$s_D = \bigl(\phi - (k/M)\,t'\bigr)(1-\gamma_Z)$. Then $\Pi$ does not
realise $\mathcal{F}^{M}_{\Lambda,0}$ with composable error
\begin{equation}\label{eq:eta-f0}
\begin{split}
  \eta \;<\; 1 &- \frac{\varepsilon_{\mathrm{sec}}}{6}
   - B_n(\theta - s;\, g_Z + s)
   - e^{-2kt'^2/(1-(k-1)/N)} \\
  &- B_M(s_D - \tau;\, g_Z + s_D)
   - B_M\bigl(\tau;\, g_Z(1-g_Z)\bigr) .
\end{split}
\end{equation}
At the operating point of \Cref{ssec:test-overhead} the best choice of
$s$, $t'$ and $\tau$ gives $\eta \geq 0.43$, and at the loose point
$\eta \geq 0.99$. As $M/k$ grows the bound approaches about
$1 - \varepsilon_{\mathrm{sec}}/3$.
\end{proposition}
\begin{proof}
Let $C^*$ draw a uniformly random set $V$ of $\phi N$ pairs and announce
$m^{(i)} = r^{(i)} \oplus 01$ on them, so that each pair in $V$ carries
$\sigma_x$ (visible in $Z$ only), and report truthfully elsewhere. Let
$\mathcal{E}$ output $1$ if the execution aborts. Otherwise it measures
every delivered pair in the $Z$ basis, computes the anti-correlated
fraction $\hat{a}_D$, and outputs $1$ if and only if
$\hat{a}_D \leq g_Z + \tau$.

\emph{Ideal execution.} At $f = 0$, $\mathcal{F}$ outputs $\bot$ unless
the delivered pairs carry no deviation, in which case it writes
$M$ independent copies of $\Lambda(\ket{\phi^+}\!\bra{\phi^+})$. Each is
anti-correlated in $Z$ with probability $g_Z$, so by~\cref{eq:conc}
\begin{equation}
\begin{split}
  \Pr[\mathcal{E}(\mathrm{ideal}) = 0]
  &\;\leq\; \Pr[\hat{a}_D > g_Z + \tau] \\
  &\;\leq\; B_M\bigl(\tau;\, g_Z(1-g_Z)\bigr),
\end{split}
\end{equation}
whatever the simulator does.

\emph{Real execution.} $\mathcal{E}$ outputs $1$ only if the protocol
aborts, or if it accepts and $\hat{a}_D \leq g_Z + \tau$. The $X$ test
sees honest pairs only, so it rejects with probability at most
$\varepsilon_{\mathrm{sec}}/6$, as in \Cref{sssec:false-rejection}. In
the $Z$ basis the batch rate is $a_Z = g_Z + s$
by~\cref{eq:per-basis-a}, and the test rejects only if $\hat{a}_Z$
exceeds it by $\theta - s$; by \Cref{lem:conc}, with the variance
bounded by $a_Z$, this has probability at most
$B_n(\theta - s;\, g_Z + s)$. For the delivered pairs, $V$ is drawn
independently of $T$, so~\cref{eq:hypergeom}, applied to the upper tail,
gives $|V \cap T| \leq k(\phi + t')$ except with probability
$e^{-2kt'^2/(1-(k-1)/N)}$. On the complement the density of $V$ in $D$
is at least $\phi - (k/M)t'$, and by (A7) the delivered outcomes are
independent with mean at least $g_Z + s_D$. The lower tail
of~\cref{eq:conc} then bounds $\Pr[\hat{a}_D \leq g_Z + \tau]$ by
$B_M(s_D - \tau;\, g_Z + s_D)$; the bound only grows as the mean falls,
so the smallest mean is the worst case. Adding the four terms bounds
$\Pr[\mathcal{E}(\mathrm{real}) = 1]$, and subtracting it from
$\Pr[\mathcal{E}(\mathrm{ideal}) = 1]$ gives~\cref{eq:eta-f0}.
\end{proof}

An honest execution realises $\mathcal{F}^{M}_{\Lambda,0}$,
with a simulator that reports truthfully, so the whole of this gap is
due to Charlie. As anticipated in \Cref{sec:nogo}, it comes from a
difference in resolution: the test cannot detect a shift much smaller
than $\theta$, while the environment, measuring all $M$ delivered
pairs, resolves shifts of order $\sqrt{\gamma_Z/M}$. At $f = f_{\mathrm{target}}$ the
functionality allows the entire window, and \Cref{thm:composable} shows that
the protocol then realises it.

\begin{theorem}\label{thm:composable}
Suppose the honest noise acts on the $i$th delivered pair as
$\Lambda_i$, and that the $\Lambda_i$ induce the
baselines $\gamma_Z, \gamma_X$ on average. Then for every dishonest Charlie $C^*$
within the threat model of \Cref{ssec:threat-scope} there is a simulator
$S$, using $C^*$ only as a black box, such that for every environment
$\mathcal{E}$
\begin{equation}
  \bigl|\Pr[\mathcal{E}(\mathrm{real}_{\Pi,C^*}) = 1]
    - \Pr[\mathcal{E}(\mathrm{ideal}_{\mathcal{F},S}) = 1]\bigr|
  \;\leq\; \tfrac{2}{3}\varepsilon_{\mathrm{sec}} ,
\end{equation}
where $\mathcal{F} = \mathcal{F}^{M}_{\Lambda,f_{\mathrm{target}}}$
and $k$ satisfies~\cref{eq:k}. The protocol of \Cref{sec:protocol}
therefore realises $\mathcal{F}^{M}_{\Lambda,f_{\mathrm{target}}}$
with composable security error at most $\tfrac{2}{3}\varepsilon_{\mathrm{sec}}$.
\end{theorem}
\begin{proof}
Under the threat model every $C^*$ factors as
\begin{equation}\label{eq:factorisation}
  C^* \;=\; \mathcal{A} \circ \mathrm{BSM}^{\otimes N},
\end{equation}
with $\mathrm{BSM}$ (Bell-state measurement) fixed and known and $\mathcal{A}$ an arbitrary
classical algorithm, unbounded, stateful and adaptive, mapping the
outcomes seen so far, together with anything the environment has sent
it, to reports and to messages back. By property~1 of
\Cref{lem:attack} the output of $\mathrm{BSM}$ is uniform and
independent across pairs, so $\mathcal{F}$ can supply $r$ at step~1 and
$S$ never reads the register holding $r$, which keeps it black-box.
Step~1 hands the ideal adversary the whole string $r$ at once, whereas a
real Charlie measures pair by pair; conceding more makes $\mathcal{F}$
harder to realise, and $S$ feeds $\mathcal{A}$ the outcomes in index
order regardless.

The simulator is the following. $S$ receives
$r^{(1)},\dots,r^{(N)}$ from $\mathcal{F}$ and runs
$\mathcal{A}$ on them in index order, relaying in both
directions between $\mathcal{A}$ and the environment. It returns to $\mathcal{F}$ whatever
report string $\mathcal{A}$ produces, or $\bot$ if $\mathcal{A}$
aborts. When $\mathcal{F}$ announces $T$, its partition and $\pi$ at
step~4, and the test outcomes at step~5, $S$ forwards them to
$\mathcal{A}$ and continues to relay. It prepares no state, measures nothing, and holds no
quantum register. We check that the two worlds agree at the honest
interface, entry by entry.

\emph{The reports.} The outcomes $r^{(i)}$ that $S$ feeds $\mathcal{A}$
are uniform and independent, as in a real execution, because
the state Charlie receives is maximally mixed. Since $\mathcal{A}$ is
the same algorithm in both worlds and sees the same distribution on its
inputs and the same environment, the reports it produces, and everything
it sends the environment, are identically distributed.

\emph{The test set and the permutation.} In the real execution Alice
draws $T$, its partition and $\pi$ under (A3); in the ideal execution
$\mathcal{F}$ draws them at step~4. Both are uniform and independent of
everything drawn before them. In both worlds the adversary learns $T$,
its partition and $\pi$, and only after his report is fixed: in the real
execution by reading the public channel of (A3), in the ideal one from
$\mathcal{F}$ at step~4, through $S$. Everything $\mathcal{A}$ sends the
environment afterwards is therefore identically distributed in the two
worlds, including any claim about where its corrupted pairs were
delivered.

\emph{The empirical fractions and the decision.} In the real execution
Alice applies $P_{m^{(i)}}$ to $A_1^{(i)}$ and the two parties measure
pair $i \in T_b$ in basis $b$. By property~3 of \Cref{lem:attack} the
correction is the last operation either performs, so what they measure
is $(P_{e_i}\otimes I)\Lambda_i(\ket{\phi^+}\!\bra{\phi^+})
(P_{e_i}\otimes I)^\dagger$ with $e_i = m^{(i)}\oplus r^{(i)}$, and by
(A7) the noise is independent across pairs. That is precisely the
distribution step~5 samples, on the same $T$ and the same $e$. The
agreement is outcome by outcome, which is what lets $S$ relay the
individual outcomes to $\mathcal{A}$. Hence $\hat{a}_Z, \hat{a}_X$
agree, and so does the acceptance decision.

\emph{The registers retained on $T$.} Consumed in both: measured in the
real execution at step~4 of the protocol, discarded by $\mathcal{F}$ at
step~5.

\emph{The delivered registers.} On acceptance the real execution
delivers, on the pairs in $D$, the state
$(E\otimes I)[\bigotimes_{i\in D}\Lambda_i(\ket{\phi^+}\!\bra{\phi^+})]
(E\otimes I)^\dagger$ in the order $\pi$, the same $\pi$ that
$\mathcal{A}$ holds in both worlds: property~3
of \Cref{lem:attack} places the relabelled correction last, so it acts
on whatever the channel left, and the noise on pair $i$ is
$\Lambda_i$ by hypothesis. This is what $\mathcal{F}$ writes
on the same input. The two agree unless $\mathcal{F}$ overrides with
$\bot$ at step~6, which it does only when $E$ has weight above
$f_{\mathrm{target}}M$, that is only on the event that the protocol
accepted with $f > f_{\mathrm{target}}$. By \Cref{thm:security} that
event has probability at most $\tfrac{2}{3}\varepsilon_{\mathrm{sec}}$.
By (A7), conditioning on acceptance, which is determined by the test
pairs and by $e$, leaves the delivered pairs' noise unchanged, so every
other entry agrees and this is the whole of the distinguishing
advantage.
\end{proof}

Now that we have a composable guarantee with noise, note that it asks
more of calibration than \Cref{thm:security}. 
A baseline off by
$\Delta$ only weakens \Cref{thm:security} to~\cref{eq:inflated}, but
\Cref{thm:composable} needs the baselines to be right. An environment
holding all $M$ delivered pairs resolves the anti-correlation rate to
$\sqrt{a(1-a)/M}$, against the protocol's $\sqrt{a(1-a)/k}$, which at the
operating point are $0.016$ and $0.038$ percentage points. The test
itself accepts any discrepancy within the $2\theta = 0.39$ percentage
points that \Cref{lem:misspec} allows, far wider than either. 
A discrepancy
the protocol accepts is therefore one the environment can still see.

\subsection{Composing with distillation}\label{ssec:distil-compose}

We now compose $\mathcal{F}^{M}_{\Lambda,f_{\mathrm{target}}}$ with
entanglement distillation. 
For this, the distiller must see a bounded per-pair error, and the
errors must be positioned as distillation's analysis assumes.

\begin{lemma}\label{lem:effective-infidelity}
Let $\Lambda$ be Bell-diagonal with weights
$(p_{\phi^+},p_{\phi^-},p_{\psi^+},p_{\psi^-})$ and
$p_{\phi^+} = 1-\delta$, and let a delivered pair additionally carry
$P_e$ with probability $f_e$, where $\sum_{e \neq 00} f_e = f$. Then the
resulting state is Bell-diagonal with fidelity
\begin{equation}\label{eq:eff-fid}
  F_{\mathrm{eff}} \;=\; \sum_{e} f_e \, p_{P_e \phi^+}
  \;\geq\; (1-f)(1-\delta),
\end{equation}
so $1 - F_{\mathrm{eff}} \leq f + \delta$, with equality approached when
Charlie places his deviations on the Bell components $\Lambda$ leaves
unpopulated.
\end{lemma}
\begin{proof}
Conjugation by $P_e$ permutes the Bell basis, so the composite is the
convolution of the two distributions over the Klein four-group, which is
the first equality. Bounding every $p_{P_e\phi^+} \geq 0$ for $e \neq
00$ and $f_{00} = 1-f$ gives the inequality; expanding
$(1-f)(1-\delta) = 1 - f - \delta + f\delta$ gives the last claim.
\end{proof}

The honest channel need not be the same on every pair
(\Cref{rem:index-noise}). If pair $i \in D$ has honest Bell weights
$\lambda_i$ with infidelity $\delta_i$ and carries $P_{e_i}$, with a
fraction $f$ of the $e_i$ non-trivial, then a delivered pair taken in
the order $\pi$ has the averaged weights
$\bar{w} = M^{-1}\sum_{i\in D}\lambda_i\circ P_{e_i}$, and the same
argument applied term by term gives
\begin{equation}\label{eq:eff-fid-avg}
  1 - F_{\mathrm{eff}}
  \;\leq\; f + \bar{\delta},
  \qquad
  \bar{\delta} = \frac{1}{M}\sum_{i \in D}\delta_i ,
\end{equation}
since $M F_{\mathrm{eff}} \geq \sum_{i:\,e_i = 00}(1-\delta_i)
\geq M(1-\bar{\delta}) - fM$. Charlie may place his misreports on the
noisiest pairs, and~\cref{eq:eff-fid-avg} holds whichever he chooses.

By \Cref{rem:symmetrisation}, the corrupted pairs are a uniformly random
$fM$-subset of the delivered ones, even conditioned on acceptance. A
distillation round consumes pairs in small groups, two at a time for
the recurrence protocols~\cite{bennett1996purification,deutsch1996quantum},
and its analysis assumes groups drawn independently. A uniformly random
subset is instead hypergeometric, hence negatively correlated, across
groups; what the composition needs is that the \emph{marginal} on the
few pairs one round consumes is close to the independent one.

\begin{lemma}\label{lem:marginal-iid}
Fix a deviation pattern $e_1,\dots,e_M$ on the delivered pairs, of
weight $w$, and let $\sigma$ be the state obtained by applying it in the
order $\pi$. (If the honest channels $\Lambda_i$ differ between pairs,
read the label of pair $i$ as $(\Lambda_i, e_i)$ throughout; the
statement and proof are unchanged.) Let $\tau$ be the state in which each delivered pair
independently carries a deviation drawn from the empirical distribution
of $e_1,\dots,e_M$, so that in particular it deviates with probability
$f = w/M$. Then for any set $S$ of $s$ delivered pairs,
\begin{equation}\label{eq:marginal-iid}
  \bigl\| \sigma|_S - \tau|_S \bigr\|_1
  \;\leq\; \frac{s(s-1)}{M} \;\leq\; \frac{s^2}{M} .
\end{equation}
\end{lemma}
\begin{proof}
Both states are classical mixtures indexed by the string of labels
appearing on $S$, over the same set of conditional states, so the trace
distance is at most $\|P-Q\|_1 = 2\,d_{\mathrm{TV}}(P,Q)$, where $P$ and
$Q$ are the two label distributions on $S$. Under $\sigma$ that string
is an ordered sample of size $s$ drawn \emph{without} replacement from
the population $\{e_1,\dots,e_M\}$, since $\pi$ is a uniform
permutation; under $\tau$ it is a sample of the same size drawn
\emph{with} replacement.

Couple the two by drawing indices $i_1,\dots,i_s$ uniformly from
$\{1,\dots,M\}$ with replacement and reading off $e_{i_1},\dots,e_{i_s}$
for $\tau$; on the event that the $i_j$ are distinct, the same indices
are distributed as a without-replacement sample and yield $\sigma$. The
two therefore differ only when some index repeats, and by a union bound
over pairs that has probability at most $\binom{s}{2}/M = s(s-1)/2M$.
The coupling inequality gives
$d_{\mathrm{TV}}(P,Q) \leq s(s-1)/2M$, and doubling
gives~\cref{eq:marginal-iid}.
\end{proof}

A recurrence protocol consuming pairs in groups of two uses $s = 2^{r}$
of them over $r$ rounds, so the error term
in~\cref{eq:marginal-iid} is $4^{r}/M$, below $10^{-3}$ at $M = 10^{6}$
for every $r \leq 4$.

Let $\mathcal{D}_r$ denote $r$ rounds of the DEJMPS recurrence
protocol~\cite{deutsch1996quantum}, applied to the delivered pairs in
the order $\pi$; testing, permuting and then distilling follows
Aschauer and Briegel~\cite{aschauer2002private}, who analyse it
asymptotically. For input Bell-diagonal weights
$p_{\phi^+}, p_{\phi^-}, p_{\psi^+}, p_{\psi^-}$, as in
\Cref{lem:delta-gamma}, one round succeeds with probability
\begin{equation}\label{eq:dejmps-succ}
  p_{\mathrm{succ}} \;=\; (p_{\phi^+} + p_{\psi^-})^2
    + (p_{\psi^+} + p_{\phi^-})^2
\end{equation}
and, conditioned on success, returns the weights
\begin{equation}\label{eq:dejmps}
\begin{aligned}
  p'_{\phi^+} &= \frac{p_{\phi^+}^2 + p_{\psi^-}^2}{p_{\mathrm{succ}}}, &
  p'_{\psi^-} &= \frac{2\,p_{\psi^+}p_{\phi^-}}{p_{\mathrm{succ}}}, \\
  p'_{\psi^+} &= \frac{p_{\psi^+}^2 + p_{\phi^-}^2}{p_{\mathrm{succ}}}, &
  p'_{\phi^-} &= \frac{2\,p_{\phi^+}p_{\psi^-}}{p_{\mathrm{succ}}} .
\end{aligned}
\end{equation}
We write $\mathrm{dej}$ for the induced map on Bell-diagonal weights,
and use only that it is fixed, that it has a fidelity threshold below
which iterating it fails to converge, and that $p_{\mathrm{succ}}$
determines the yield.

\begin{proposition}\label{prop:distil-compose}
Assume the hypotheses of \Cref{thm:composable}, and let $\lambda_i$ be
the Bell weights of the honest channel $\Lambda_i$ on delivered pair
$i$. Let
\begin{equation}\label{eq:admissible}
  \mathcal{W} \;=\; \Bigl\{\, \frac{1}{M}\sum_{i\in D}
  \lambda_i \circ P_{e_i} \;:\;
  \bigl|\{i : e_i \neq 00\}\bigr| \leq f_{\mathrm{target}} M \,\Bigr\}
\end{equation}
be the set of averaged Bell weights that a pattern the functionality
accepts can produce, and suppose every $w \in \mathcal{W}$ lies in the
basin on which iterating $\mathrm{dej}$ converges. For $w \in
\mathcal{W}$ write $w^{(j)} = \mathrm{dej}^{j}(w)$, $F^{(r)}(w)$ for the
fidelity of $w^{(r)}$, and
$p_{\mathrm{succ},j}(w) = p_{\mathrm{succ}}\bigl(w^{(j-1)}\bigr)$ for the
success probability~\cref{eq:dejmps-succ} assigns to round $j$. Then running
$\mathcal{D}_r$ on the output of
$\mathcal{F}^{M}_{\Lambda,f_{\mathrm{target}}}$ returns pairs of fidelity
at least
\begin{equation}\label{eq:dej-worst}
  F^{(r)}_{\min} \;=\; \min_{w \in \mathcal{W}} F^{(r)}(w),
\end{equation}
in expected number at least
\begin{equation}\label{eq:dej-yield}
  \frac{M}{2^{r}}\,\min_{w \in \mathcal{W}}\prod_{j=1}^{r} p_{\mathrm{succ},j}(w) ,
\end{equation}
up to an error $O(4^{r}/M)$ from \Cref{lem:marginal-iid}. Consequently,
running $\mathcal{D}_r$ on the output of the protocol of
\Cref{sec:protocol} returns pairs of fidelity at least
$F^{(r)}_{\min}$ except with probability at most
$\tfrac{2}{3}\varepsilon_{\mathrm{sec}} + O(4^{r}/M)$.
\end{proposition}
\begin{proof}
On acceptance the functionality delivers
$(E\otimes I)[\bigotimes_{i\in D}\Lambda_i(\ket{\phi^+}\!\bra{\phi^+})]
(E\otimes I)^\dagger$ with $E$ of weight at most
$f_{\mathrm{target}}M$, in the order $\pi$. By \Cref{lem:marginal-iid},
with labels $(\Lambda_i,e_i)$, the pairs each round consumes are within
$O(4^r/M)$ of independent copies of the Bell-diagonal pair with the
averaged weights $w = M^{-1}\sum_{i}\lambda_i\circ P_{e_i}$, which lies
in $\mathcal{W}$ by the weight constraint. On independent copies the
recursion~\cref{eq:dejmps} applies~\cite{deutsch1996quantum}, so after
$r$ rounds the surviving pairs have weights $\mathrm{dej}^r(w)$. Charlie
chooses the pattern, and with it $w$, so the guarantee is the minimum
over $\mathcal{W}$. For the
yield~\cref{eq:dej-yield}: each output of a round consumes two of its
inputs, giving the factor $2^{-r}$, and a round is attempted only on
pairs its predecessors returned, so the success probabilities multiply.
The second claim follows from the first by
\Cref{thm:composable}: $\Pi$ realises
$\mathcal{F}^{M}_{\Lambda,f_{\mathrm{target}}}$ with composable error
$\tfrac{2}{3}\varepsilon_{\mathrm{sec}}$, and a distinguisher for the
composed protocol is a distinguisher for the resource it
consumes.
\end{proof}

The guarantee is a minimum over $\mathcal{W}$ rather than a function of
the fidelity, because $\mathrm{dej}$ acts on the full weight vector and,
by \Cref{lem:delta-gamma}, $F_{\mathrm{eff}}$ does not determine
$(p_{\phi^+},p_{\phi^-},p_{\psi^+},p_{\psi^-})$: \cref{eq:eff-fid-avg} constrains where $w$ may sit, and
Charlie's allocation $(f_x,f_z,f_y)$, and his choice of which pairs to
corrupt, move it within that constraint. When every pair has the same
honest weights $\lambda$, $\mathcal{W}$ is contained in the tetrahedron
with vertices $\lambda$ and
$(1-f_{\mathrm{target}})\lambda + f_{\mathrm{target}}\,\lambda\circ P_e$
for $e \in \{01,10,11\}$, so~\cref{eq:dej-worst} is a minimisation over
three parameters; since $\mathrm{dej}^r$ is nonlinear the minimum need not
lie at a vertex. Evaluating it needs the honest weights $\lambda$, a
channel model beyond the two baselines the protocol measures, which is
why we do not instantiate~\cref{eq:dejmps} at the reference operating
point. Composition reaches fidelities that \Cref{sec:resources} shows
verification alone cannot: verify at an affordable
$f_{\mathrm{target}}$, then distil to the fidelity the application
needs. Verification adds $k$ pairs, while distillation multiplies the
number required by $2^{r}$.

\section{Resource estimation} \label{sec:resources}

To obtain $M$ verified EPR pairs, the protocol consumes $k$ test pairs
and requires a quantum memory of sufficient size and coherence time.

\subsection{Target misreport rate}\label{ssec:target}

The protocol does not have to rule out every misreported pair, only
enough of them that error correction can absorb the rest. That
requirement sets $f_{\mathrm{target}}$, the largest fraction of
delivered pairs that the security guarantee allows Charlie to misreport
(\Cref{def:prob-security}).

The errors that survive verification are Pauli errors on the delivered
pairs, of bounded rate and in uniformly random positions
(\Cref{cor:state,rem:symmetrisation}), which is the kind of error
quantum error correction is designed to handle, provided the rate is low
enough. The surface code tolerates a physical error rate of
order $1\%$~\cite{fowler2012surface}, and practical operation targets
rates well below threshold. An $f_{\mathrm{target}}$ of $10^{-1}$
tolerates $10^5$ corrupted pairs in $10^6$, which no fault-tolerant
architecture would accept. Moreover, \Cref{cor:state} bounds the
deviation relative to $\rho^{\mathrm{hon}}_D$, which already carries the
channel's errors at rate $\delta$, so the error-correction layer sees a
rate of order $f_{\mathrm{target}} + \delta$ (\Cref{lem:delta-gamma}); on
hardware with a minimum noise level of about one per cent, this leaves little
room for adversarial errors. We therefore take
$f_{\mathrm{target}} = 10^{-2}$ as the largest rate plausibly compatible
with fault-tolerant operation, and use it as the operating point for the
estimates that follow.

\subsection{Test pair overhead}\label{ssec:test-overhead}

\begin{figure}[h]
\centering
\begin{tikzpicture}
\begin{axis}[
  width=\columnwidth, height=0.78\columnwidth,
  xlabel={$f_{\mathrm{target}}$}, ylabel={test pairs $k$},
  xmode=log, ymode=log,
  xmin=0.0014, xmax=0.32, ymin=200, ymax=1.2e7,
  xtick={0.002,0.005,0.01,0.02,0.05,0.1,0.2,0.3},
  xticklabels={$0.002$,$0.005$,$0.01$,$0.02$,$0.05$,$0.1$,$0.2$,$0.3$},
  ytick={1e3,1e4,1e5,1e6,1e7},
  legend pos=north east, legend cell align=left,
  legend style={font=\footnotesize, fill=white, fill opacity=0.92,
                text opacity=1, draw=gray!40},
  every axis plot/.append style={very thick},
  label style={font=\footnotesize}, tick label style={font=\footnotesize},
  clip mode=individual,
]

\addplot[draw=none, fill=black, fill opacity=0.055, forget plot]
  coordinates {(0.0014,1e6) (0.32,1e6) (0.32,1.2e7) (0.0014,1.2e7)}
  \closedcycle;
\addplot[black, thin, densely dashdotted, forget plot]
  coordinates {(0.0014,1e6) (0.32,1e6)};

\addplot[blue!70!black] coordinates {
  (0.00193,162244) (0.00204,100328) (0.00216,81906) (0.00229,70052) (0.00242,61872) (0.00256,55316)
  (0.00271,49894) (0.00287,45316) (0.00304,41390) (0.00321,38154) (0.00340,35136) (0.00360,32476)
  (0.00381,30112) (0.00403,28000) (0.00427,26028) (0.00452,24262) (0.00478,22676) (0.00506,21192)
  (0.00535,19854) (0.00567,18568) (0.00600,17410) (0.00635,16334) (0.00672,15336) (0.00711,14410)
  (0.00752,13554) (0.00796,12742) (0.00843,11978) (0.00892,11276) (0.00944,10616) (0.00999,9998)
  (0.01057,9420) (0.01119,8874) (0.01184,8366) (0.01254,7880) (0.01327,7430) (0.01404,7010)
  (0.01486,6610) (0.01573,6236) (0.01664,5886) (0.01761,5554) (0.01864,5240) (0.01973,4946)
  (0.02088,4668) (0.02210,4406) (0.02339,4160) (0.02475,3928) (0.02620,3708) (0.02773,3500)
  (0.02934,3306) (0.03105,3122) (0.03287,2948) (0.03478,2786) (0.03681,2630) (0.03896,2484)
  (0.04123,2346) (0.04364,2216) (0.04618,2094) (0.04888,1978) (0.05173,1868) (0.05475,1764)
  (0.05794,1668) (0.06132,1576) (0.06490,1488) (0.06869,1406) (0.07269,1328) (0.07693,1254)
  (0.08142,1186) (0.08617,1120) (0.09120,1058) (0.09652,1000) (0.10215,944) (0.10811,892)
  (0.11442,844) (0.12109,798) (0.12816,754) (0.13564,712) (0.14355,672) (0.15192,636)
  (0.16079,600) (0.17017,568) (0.18009,536) (0.19060,506) (0.20172,478) (0.21349,452)
  (0.22594,428) (0.23912,404) (0.25307,382) (0.26784,360) (0.28346,340) (0.30000,322)
};
\addlegendentry{$\gamma_{\max}=0$}
\addplot[orange!85!black,dashed] coordinates {
  (0.00399,2724904) (0.00419,603634) (0.00441,386674) (0.00463,291636) (0.00487,231482) (0.00512,190746)
  (0.00538,161072) (0.00565,138404) (0.00594,119948) (0.00624,105142) (0.00656,92666) (0.00689,82352)
  (0.00724,73470) (0.00761,65786) (0.00800,59106) (0.00841,53278) (0.00883,48282) (0.00928,43776)
  (0.00976,39726) (0.01025,36224) (0.01078,33004) (0.01132,30204) (0.01190,27630) (0.01251,25312)
  (0.01314,23254) (0.01381,21366) (0.01451,19664) (0.01525,18110) (0.01603,16690) (0.01685,15398)
  (0.01770,14234) (0.01860,13160) (0.01955,12176) (0.02055,11272) (0.02159,10452) (0.02269,9694)
  (0.02385,8994) (0.02506,8356) (0.02634,7764) (0.02768,7222) (0.02909,6720) (0.03057,6258)
  (0.03212,5834) (0.03376,5438) (0.03548,5072) (0.03728,4734) (0.03918,4422) (0.04117,4132)
  (0.04327,3862) (0.04547,3612) (0.04779,3380) (0.05022,3164) (0.05277,2964) (0.05546,2776)
  (0.05828,2602) (0.06125,2440) (0.06437,2290) (0.06765,2148) (0.07109,2018) (0.07471,1896)
  (0.07851,1780) (0.08251,1674) (0.08671,1574) (0.09112,1480) (0.09576,1394) (0.10063,1312)
  (0.10576,1234) (0.11114,1162) (0.11680,1096) (0.12274,1032) (0.12899,974) (0.13556,918)
  (0.14246,866) (0.14971,816) (0.15733,772) (0.16534,728) (0.17375,688) (0.18260,650)
  (0.19189,614) (0.20166,580) (0.21193,548) (0.22271,518) (0.23405,490) (0.24596,464)
  (0.25848,438) (0.27164,414) (0.28547,392) (0.30000,372)
};
\addlegendentry{$\gamma_{\max}=0.01$}
\addplot[red!75!black,dotted] coordinates {
  (0.00459,11979292) (0.00482,6632526) (0.00506,3990320) (0.00532,2528892) (0.00559,1714140) (0.00587,1232624)
  (0.00616,930768) (0.00647,725130) (0.00680,579744) (0.00714,475938) (0.00750,396936) (0.00788,335412)
  (0.00828,286552) (0.00869,247930) (0.00913,215442) (0.00959,188524) (0.01008,165566) (0.01059,146254)
  (0.01112,129868) (0.01168,115628) (0.01227,103222) (0.01289,92388) (0.01354,82906) (0.01422,74582)
  (0.01494,67166) (0.01569,60642) (0.01648,54822) (0.01731,49628) (0.01818,44994) (0.01910,40816)
  (0.02006,37092) (0.02107,33736) (0.02214,30690) (0.02325,27980) (0.02442,25522) (0.02565,23296)
  (0.02695,21272) (0.02831,19444) (0.02973,17798) (0.03123,16294) (0.03281,14922) (0.03446,13684)
  (0.03620,12552) (0.03802,11526) (0.03994,10586) (0.04195,9732) (0.04407,8950) (0.04629,8236)
  (0.04862,7586) (0.05107,6990) (0.05365,6444) (0.05635,5946) (0.05919,5488) (0.06218,5068)
  (0.06531,4684) (0.06860,4330) (0.07206,4006) (0.07569,3708) (0.07951,3434) (0.08352,3182)
  (0.08773,2952) (0.09215,2738) (0.09680,2542) (0.10168,2360) (0.10680,2194) (0.11219,2038)
  (0.11784,1896) (0.12378,1766) (0.13002,1644) (0.13658,1532) (0.14346,1428) (0.15069,1332)
  (0.15829,1242) (0.16627,1160) (0.17465,1082) (0.18346,1012) (0.19270,946) (0.20242,884)
  (0.21262,828) (0.22334,774) (0.23460,726) (0.24643,680) (0.25885,636) (0.27190,596)
  (0.28560,560) (0.30000,526)
};
\addlegendentry{$\gamma_{\max}=0.05$}
\addlegendimage{area legend, draw=black, thin, densely dashdotted,
  fill=black, fill opacity=0.055}
\addlegendentry{$k>M$}
\end{axis}
\end{tikzpicture}
\caption{Test pairs required against $f_{\mathrm{target}}$,
from~\cref{eq:k}, at $\varepsilon_{\mathrm{sec}}=10^{-2}$ and $M=10^6$,
for three declared noise baselines.}
\label{fig:k-vs-f}
\end{figure}
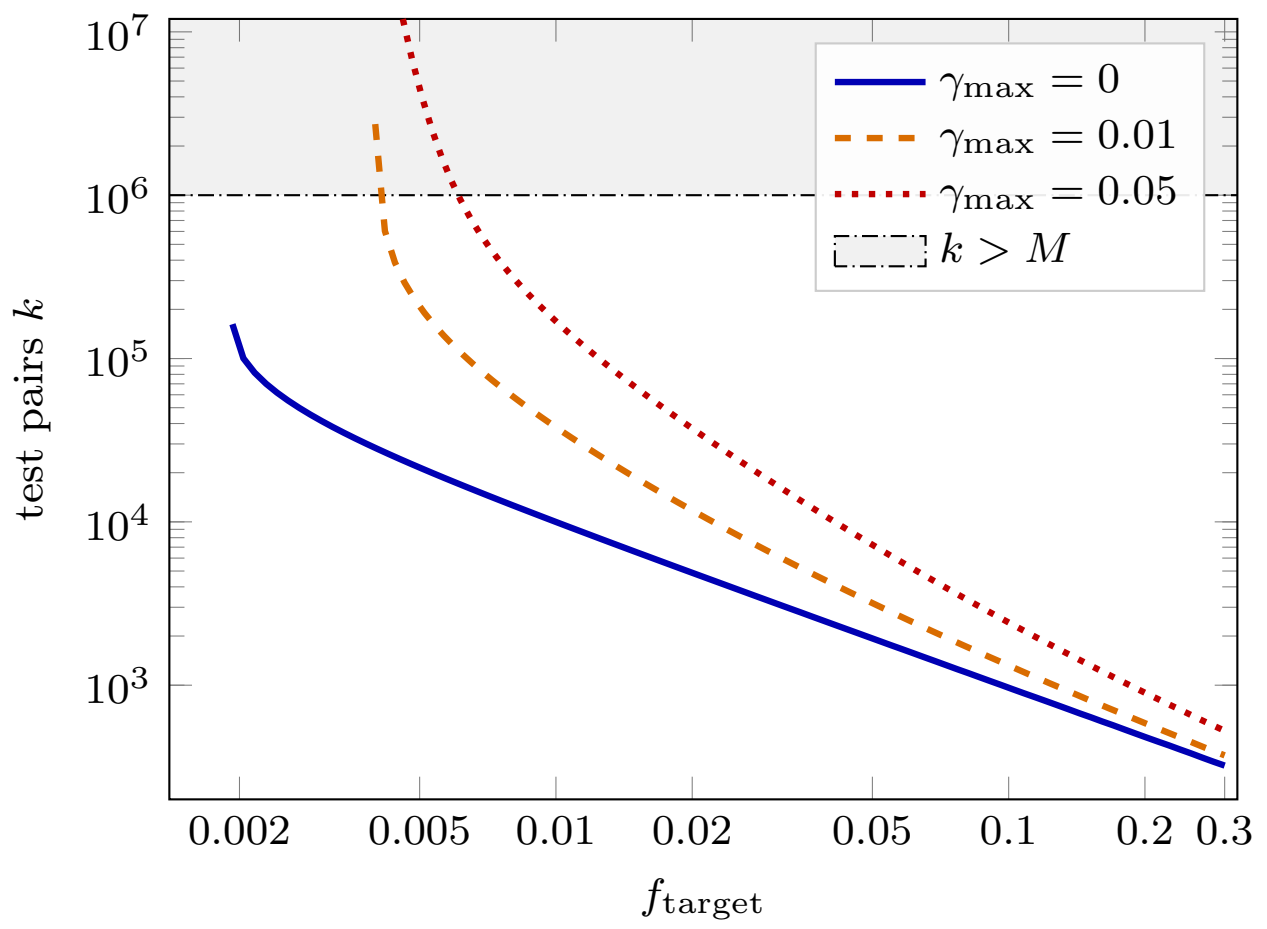

\Cref{tab:k-vs-f} and \Cref{fig:k-vs-f} give the $k$ required by
\Cref{thm:security}. In line with~\cref{eq:k-scaling}, halving
$f_{\mathrm{target}}$ costs a factor of two on hardware clean enough
that $\gamma_{\max} \lesssim f_{\mathrm{target}}$ and a factor of four
once the baseline dominates, while the dependence on
$\varepsilon_{\mathrm{sec}}$ is mild: at $f_{\mathrm{target}} = 0.1$ and
$\gamma_{\max} = 0.05$, reducing $\varepsilon_{\mathrm{sec}}$ from
$10^{-2}$ to $10^{-9}$ multiplies $k$ by about three and a half.

Two parameter points recur, both at $\varepsilon_{\mathrm{sec}} = 0.01$,
$\gamma_{\max} = 0.05$ and $M = 10^6$. The \emph{loose} point,
$f_{\mathrm{target}} = 0.1$, needs $k = 2{,}420$ test pairs, an overhead
of $0.24\%$, but its guarantee is too weak for fault-tolerant use.
The \emph{operating} point, $f_{\mathrm{target}} = 10^{-2}$, needs
$k = 168{,}982$, or $17\%$ of the delivered batch and $14\%$ of the
pairs prepared; these are the reference parameters used throughout. At
fixed $f_{\mathrm{target}}$ the fraction $k/N$ decreases as $M$ grows.

\begin{table*}[t]
\caption{Test pairs required against $f_{\mathrm{target}}$ and the
declared noise baseline, at $\varepsilon_{\mathrm{sec}} = 0.01$ and
$M = 10^6$. The last column gives $k/M$ at $\gamma_{\max} = 0.05$.}\label{tab:k-vs-f}
\begin{ruledtabular}
\begin{tabular}{lrrrr}
& \multicolumn{3}{c}{$k$ from~\cref{eq:k}} & \\
$f_{\mathrm{target}}$ & $\gamma_{\max}=0$ & $\gamma_{\max}=0.01$
  & $\gamma_{\max}=0.05$ & $k/M$ at $0.05$ \\
\hline
$0.300$ & $322$ & $372$ & $526$ & $0.1\%$ \\
$0.200$ & $482$ & $586$ & $898$ & $0.1\%$ \\
$0.150$ & $644$ & $816$ & $1{,}340$ & $0.1\%$ \\
$0.100$ & $966$ & $1{,}322$ & $2{,}420$ & $0.2\%$ \\
$0.070$ & $1{,}380$ & $2{,}058$ & $4{,}194$ & $0.4\%$ \\
$0.050$ & $1{,}934$ & $3{,}182$ & $7{,}240$ & $0.7\%$ \\
$0.040$ & $2{,}420$ & $4{,}298$ & $10{,}560$ & $1.1\%$ \\
$0.030$ & $3{,}234$ & $6{,}430$ & $17{,}510$ & $1.8\%$ \\
$0.020$ & $4{,}878$ & $11{,}754$ & $37{,}308$ & $3.7\%$ \\
$0.015$ & $6{,}548$ & $18{,}610$ & $66{,}602$ & $6.7\%$ \\
$0.010$ & $9{,}986$ & $37{,}940$ & $168{,}982$ & $16.9\%$ \\
$0.007$ & $14{,}660$ & $79{,}362$ & $514{,}428$ & $51.4\%$ \\
$0.005$ & $21{,}492$ & $208{,}354$ & $4{,}491{,}946$ & $449.2\%$ \\

\end{tabular}
\end{ruledtabular}
\end{table*}

The test budget exceeds the delivered batch below
$f_{\mathrm{target}} \approx 0.41\%$ at $\gamma_{\max} = 0.01$ and
$0.61\%$ at $\gamma_{\max} = 0.05$; on a clean channel it never does.
The left-hand side of~\cref{eq:k} has a strictly positive minimum over
$k$, so below some $f_{\mathrm{target}}$ no amount of testing lets
\Cref{thm:security} certify, which we expected. 
Once the baseline dominates, the minimum
recedes to large $k$ and the floor approaches what the sampling transfer
alone contributes,
\begin{equation}\label{eq:asymptote}
  \lim_{k \to \infty}\;\frac{2k}{M}
  \sqrt{\Bigl(1-\frac{k-1}{N}\Bigr)
   \frac{\log(6/\varepsilon_{\mathrm{sec}})}{2k}}
  \;=\; \sqrt{\frac{2\log(6/\varepsilon_{\mathrm{sec}})}{M}} ,
\end{equation}
which is $0.36\%$ at $\varepsilon_{\mathrm{sec}} = 0.01$ and
$M = 10^6$. On a clean channel the minimum is interior and the floor
drops to $0.19\%$. The three baselines of \Cref{tab:k-vs-f} give floors
of $0.19\%$, $0.36\%$ and $0.38\%$. None is intrinsic, since~\cref{eq:k}
is sufficient and not necessary, and all relax as $M^{-1/2}$: reaching
$f_{\mathrm{target}} = 10^{-3}$ would require a batch of about
$3\times10^6$ to $4\times10^7$ delivered pairs, depending on the
baseline.

The floor is set by the sampling transfer, which can be bounded in three
ways:
\begin{enumerate}
 \item crudely, by $f \leq (N/M)f_N$;
 \item by keeping the deviations caught in the sample,
 as~\cref{eq:transfer} does, but with plain Hoeffding;
 \item with Serfling's correction, as~\cref{eq:hypergeom} does.
\end{enumerate}
At $\gamma_{\max}=0.05$ and $f_{\mathrm{target}}=0.01$ they ask for
$169{,}714$, $174{,}402$ and $168{,}982$ test pairs respectively. What
separates them is reach: the first two admit no solution below
$f_{\mathrm{target}} = 0.68\%$ and $0.72\%$, and Serfling's correction
moves that floor to $0.38\%$. A Bernstein form of Serfling's
bound~\cite{bardenet2015concentration} could lower the floor
further.

The variance-aware bound of~\cref{eq:conc} matters most for the
estimation margins. A margin from Hoeffding's inequality would need
$2.4\times10^6$ test pairs at the operating point, about fourteen times the
$168{,}982$ above. The cost is in calibration: at the operating point
$\theta = 0.19\%$, so a $\gamma_b$ declared a fifth of a percentage point
from the truth either costs liveness or, in the direction of
\Cref{rem:overest}, allows as much undetected deviation as the target
was meant to exclude. The sharper bound thus trades test pairs for
calibration accuracy, and makes the trade-off explicit.

For comparison, verifying a pure entangled state to infidelity
$\delta$ needs $\Theta(\delta^{-1}\log(1/\varepsilon_{\mathrm{sec}}))$
samples, locally~\cite{pallister2018optimal} and
adversarially~\cite{zhu2019efficient} (reviewed in~\cite{yu2022statistical}).
Our $k$ is
larger by $(\gamma_{\max}+f_{\mathrm{target}})/f_{\mathrm{target}}$,
because $f_{\mathrm{target}}$ must be resolved against a minimum noise level
present whether or not anyone cheats. At $f_{\mathrm{target}} = 10^{-2}$
the test costs $1\%$, $3.8\%$ and $17\%$ of the delivered batch at
baselines of $0$, $1\%$ and $5\%$, none prohibitive; the limiting
constraints are memory and calibration.

\subsection{Quantum memory requirement}\label{ssec:memory}

The test-pair count is cheap. What makes the protocol hard to run is
when the test pairs are chosen. By (A3), Alice picks them only after
Charlie has announced every report, which stops him from aiming his
deviations at the pairs that will not be tested. Until that point any
pair might be tested, so Alice and Bob cannot measure or use any of them.
They must store all $N$ pairs until the last report arrives, and this
sets how long the memory has to last and how many qubits it needs.

Holding every pair until all $N$ reports have been received needs a
coherence time of at least
\begin{equation}\label{eq:coherence}
  T_{\mathrm{coh}} \geq N \cdot \tau_{\mathrm{swap}},
\end{equation}
where $\tau_{\mathrm{swap}}$ is the time per delivered swap round. The
requirement scales with the entire batch, not with a link round-trip.
Heralded entanglement between matter qubits is delivered at $9$ and
$7\,\mathrm{Hz}$ on the two links of the three-node network of
Pompili et al.~\cite{pompili2021realization}, whose nodes are
$2$--$7\,\mathrm{m}$ apart. At $N = 10^6$, \cref{eq:coherence} then asks
for $T_{\mathrm{coh}}$ of the order of a day. This is only a necessary
condition. A memory that barely outlasts the batch adds its own noise to
the oldest pairs, and once this is counted it must last $40$ to $50$ times
longer than the batch (\Cref{ssec:lifetime}). Demonstrated memories fall
far short of this. In the same network, the memory qubit decays under
continued entanglement generation with a constant of
$N_{1/e} = 1843 \pm 32$ attempts. Each attempt lasts
$3.8\,\mu\mathrm{s}$ and succeeds with probability between
$3.4\times10^{-5}$, implied by the $9\,\mathrm{Hz}$ rate, and
$4.7\times10^{-5}$, the heralding probability computed from the emission
and detection parameters reported there, so $N_{1/e}$ corresponds to
only $0.06$--$0.09$ heralded pairs. The memory's intrinsic dephasing
time, measured without entanglement attempts, is $11.6\,\mathrm{ms}$,
likewise only about a tenth of the $110\,\mathrm{ms}$ between heralded
pairs. A ten-qubit solid-state register, left idle, preserves a single-qubit state for
over $75\,\mathrm{s}$ and two-qubit entanglement for over
$10\,\mathrm{s}$~\cite{bradley2019ten}, some $675$ and $90$ pairs at
$9\,\mathrm{Hz}$. Repeater networks already require long-lived
memories~\cite{wehner2018quantum}, but only for establishing and
confirming a single link, so~\cref{eq:coherence} is an additional
requirement.

The second cost is capacity. The $N$ pairs are held \emph{at once}, so
each party needs a quantum memory of $N = M+k$ addressable qubits rather
than a single register reused $N$ times: a million qubits per end node at
$M = 10^6$, against one memory qubit per node in the network of
Pompili et al.~\cite{pompili2021realization} and ten qubits in the
register of Bradley et al.~\cite{bradley2019ten}. The two costs are
independent: a memory with unbounded coherence but a handful of qubits
serves no better than a million short-lived ones.

Three modifications relax~\cref{eq:coherence}. The most effective, and
the most practical, splits the batch into blocks, each verified before
the next begins; if each block is released as soon as it is verified,
coherence and capacity fall to the block size and the number of
delivered pairs can grow without limit (\Cref{ssec:blocks}). The other
two remove the
test pairs from~\cref{eq:coherence}, so that each can be measured as it
arrives: Charlie commits to each outcome as he obtains it and opens the
commitments once the batch closes, at the price of a computational
assumption~\cite{naor1991bit}; or $T$ is drawn and kept secret in advance, at the cost of a
pre-shared secret key linear in $N$.

\subsection{Test pairs and memory together}\label{ssec:joint}

The two sections above pull in opposite directions. Certifying a
tighter bound (a smaller $f_{\mathrm{target}}$) takes more pairs, both
more delivered pairs $M$ and more test pairs $k$. But every pair has to
sit in memory until the batch closes, so the memory wants as few pairs
as possible. In practice the memory is the hard limit. The useful
question is therefore: for a memory that holds $N$ pairs, how should
they be divided between delivered and test pairs?
\Cref{tab:joint} answers this. For each budget $N$ it gives the split
$N = M + k$ that certifies the smallest $f_{\mathrm{target}}$ allowed
by~\cref{eq:k}, and the memory lifetime that split requires.

\begin{table*}[t]
\caption{Best certified rate and required memory lifetime against
memory budget $N$, at $\varepsilon_{\mathrm{sec}} = 0.01$. The split
$(M,k)$ is optimal at $\gamma_{\max} = 0.05$; the lifetime, from~\cref{eq:lifetime},
keeps the oldest pair within $\gamma_{\mathrm{cap}} = 0.05$ for
preparation noise $\gamma_0$. Times are at the $9\,\mathrm{Hz}$ link
rate of~\cite{pompili2021realization}.}\label{tab:joint}
\begin{ruledtabular}
\begin{tabular}{rrrrrrrr}
& & \multicolumn{2}{c}{best $f_{\mathrm{target}}$} & &
  \multicolumn{2}{c}{lifetime $T_{\mathrm{mem}}/\tau_{\mathrm{swap}}$} & lifetime \\
$N$ & batch duration & $\gamma_{\max}=0$ & $\gamma_{\max}=0.05$
  & split $(M,k)$ & $\gamma_0 = 0$ & $\gamma_0 = 0.01$
  & at $9\,\mathrm{Hz}$ \\
\hline
$10^3$ & $2\,\mathrm{min}$ & $30\%$ & $42\%$ & $(380,\,620)$
  & $3.9\times10^{4}$ & $4.8\times10^{4}$ & $1.5\,\mathrm{h}$ \\
$10^4$ & $19\,\mathrm{min}$ & $5.3\%$ & $10\%$ & $(4930,\,5070)$
  & $3.9\times10^{5}$ & $4.8\times10^{5}$ & $15\,\mathrm{h}$ \\
$10^5$ & $3.1\,\mathrm{h}$ & $1.0\%$ & $2.8\%$ & $(55534,\,44466)$
  & $3.9\times10^{6}$ & $4.8\times10^{6}$ & $6.2\,\mathrm{d}$ \\
$7.3\times10^5$ & $23\,\mathrm{h}$ & $0.26\%$ & $1.0\%$
  & $(422280,\,307720)$
  & $2.8\times10^{7}$ & $3.5\times10^{7}$ & $46\,\mathrm{d}$ \\
\end{tabular}
\end{ruledtabular}
\end{table*}

No demonstrated memory reaches even the first row. Certifying any rate
below one requires $N \geq 220$ on a clean channel and $N \geq 285$ at a
five-per-cent baseline, and a rate of one half requires $519$ and $761$,
whereas the networked memory of Pompili et al.\ holds about $0.1$
heralded pairs in a single qubit, and the idle ten-qubit register is
limited by its capacity to ten. Reaching $f_{\mathrm{target}} = 10^{-2}$
needs $N \geq 1.05\times10^5$ on a clean channel and $7.3\times10^5$ at a
five-per-cent baseline; at $N = 7.3\times10^5$ the optimal split is
$(422280,\,307720)$ at that baseline and $(630194,\,99806)$ on a clean
channel. The gap is one of memory hardware. The certified rate improves
monotonically with the budget, but both capacity and coherence must grow by several orders of magnitude
before the protocol certifies a useful rate.

\subsection{Memory lifetime under storage noise}\label{ssec:lifetime}

The estimates so far treat the baselines as fixed. But the memory
holding the batch adds noise of its own, and by \Cref{rem:index-noise}
the baseline that enters \Cref{thm:security} is the per-pair maximum,
since Charlie can choose which pairs to corrupt.
Pair $i$ is held from its preparation until the test set is drawn,
about $(N-i)\,\tau_{\mathrm{swap}}$, so the oldest pair is held for
$N\tau_{\mathrm{swap}}$. Suppose each stored qubit loses coherence with
time constant $T_{\mathrm{mem}}$, by depolarising or by dephasing. The
correlation of a stored pair in the noisier basis then decays as
$e^{-2t/T_{\mathrm{mem}}}$, one factor for each stored half, and a pair
with baseline $\gamma_0$ at preparation has, after storage time $t$,
\begin{equation}\label{eq:gamma-t}
  \gamma(t) \;=\; 1 - (1-\gamma_0)\,e^{-2t/T_{\mathrm{mem}}} .
\end{equation}
Requiring the oldest pair to stay within a ceiling
$\gamma_{\max} \leq \gamma_{\mathrm{cap}}$ fixes
\begin{equation}\label{eq:lifetime}
  \frac{T_{\mathrm{mem}}}{\tau_{\mathrm{swap}}}
  \;\geq\; \frac{2N}{\log\bigl[(1-\gamma_0)/(1-\gamma_{\mathrm{cap}})\bigr]}
  \;\approx\; \frac{2N}{\gamma_{\mathrm{cap}} - \gamma_0} .
\end{equation}
So the memory must last about $2/(\gamma_{\mathrm{cap}} - \gamma_0)$
times longer than the batch, where \cref{eq:coherence} alone asks for
one batch length. At $\gamma_{\mathrm{cap}} = 5\%$ this is $39$ times
for pairs prepared perfectly, and $48$ times for pairs prepared with
$\gamma_0 = 1\%$.

The last three columns of \Cref{tab:joint} apply~\cref{eq:lifetime} to
each budget. Certifying $f_{\mathrm{target}} = 1\%$ requires each
memory qubit to last $2.8$--$3.5\times10^{7}$ swap rounds, five to seven
weeks at $9\,\mathrm{Hz}$. Even the smallest budget, which certifies
only $42\%$, requires $4$--$5\times10^{4}$ rounds, about an hour and a
half, some sixty times the idle lifetime of about $675$ rounds of the
ten-qubit register~\cite{bradley2019ten} and five orders of magnitude beyond
the networked one~\cite{pompili2021realization}.

Raising the ceiling helps little. Optimising jointly over $N$, the split
and the ceiling, at $\gamma_0 = 1\%$, moves the optimum to
$\gamma_{\max} \approx 0.22$ and lowers the lifetime needed for $1\%$
only to $2.3\times10^{7}$ rounds. It also leaves the oldest delivered
pairs with an infidelity of at least $6\%$ by
\Cref{lem:delta-gamma}, too noisy for the error-correction layer of
\Cref{ssec:target}. The same factor $2/(\gamma_{\mathrm{cap}}-\gamma_0)$
applies to the per-block release of \Cref{tab:blocks}, with $N$ replaced
by $n_b$.

\subsection{Block-wise verification}\label{ssec:blocks}

Splitting the batch into $B$ blocks of $n_b = N/B$ pairs, each verified
before the next begins, removes the need to hold every pair through the
whole batch. Charlie then sees the test sets and outcomes of earlier
blocks before committing to later ones, but this adaptivity does not
weaken the sampling bounds (\Cref{lem:pooled}, Appendix~\ref{app:blocks}).

How much memory blocking saves depends on when the delivered pairs are
released. Under \emph{end-of-batch release} the decision is taken once,
on the pooled statistic. Each block's test pairs need survive only $n_b$
rounds, and capacity falls from $M+k$ to $M+k_b$, but the delivered pairs
are held until the decision, so~\cref{eq:coherence} stands for them
unchanged. Under \emph{per-block release} block $j$'s delivered pairs
pass to the application as block $j$ is verified, and coherence and
capacity both fall to $n_b$. But a guarantee on those pairs can rest only
on evidence available when they are released, so each block must certify
itself, and there is nothing left to pool. The delivery permutation
$\pi$ of step~5 is then drawn within each block, after that block's
reports and independently of its test set, so
\Cref{rem:symmetrisation} holds block by block. The two benefits cannot
be combined: pooling is available only when the coherence saving is not.

Per-block release is nonetheless useful. A block of $n_b$ pairs
certifies a rate set by $n_b$, and the number of blocks can then grow
without limit, at the cost only of the $\log B$ that a union bound over
blocks adds (\Cref{tab:blocks}). A memory of $10^4$ qubits, for example,
certifies $10\%$ on a single block and $13.8\%$ on $4.7\times10^5$
delivered pairs over $100$ blocks; blocks of $10^5$ bring the rate down
to $3$--$4\%$. Blocking does not lower the memory a given rate needs,
which is close to that of a single batch of the same size
(\Cref{tab:joint}), but it does remove the link between the memory and
the number of pairs delivered.

\begin{table}[h]
\caption{Certified rate and delivered pairs under per-block release,
at $\gamma_{\max} = 0.05$ and $\varepsilon_{\mathrm{sec}} = 0.01$, with
coherence and capacity both held to the block size $n_b$.}\label{tab:blocks}
\begin{ruledtabular}
\begin{tabular}{rrrr}
block size $n_b$ & blocks $B$ & delivered $M$ & $f_{\mathrm{target}}$ \\
\hline
$10^3$    & $1$    & $380$             & $41.6\%$ \\
$10^3$    & $100$  & $34{,}800$        & $60.1\%$ \\
$10^3$    & $1000$ & $336{,}000$       & $68.6\%$ \\
$10^4$    & $1$    & $4{,}930$         & $10.0\%$ \\
$10^4$    & $100$  & $470{,}800$       & $13.8\%$ \\
$10^5$    & $10$   & $549{,}700$       & $3.3\%$  \\
$10^5$    & $100$  & $5.45\times10^6$  & $3.8\%$  \\

\end{tabular}
\end{ruledtabular}
\end{table}

Per-block release admits pairs to the application before the batch is
complete, which \Cref{rem:isolation} cautions against, so it suits a
deployment that consumes pairs as they arrive. Moreover, for every
block size shown the certified rate in \Cref{tab:blocks} remains above
the target of \Cref{ssec:target}, and even the smallest block is far
beyond demonstrated memories. Blocking makes the achievable rate depend
on block size alone.

\section{Discussion}\label{sec:discussion}

We studied entanglement swapping through a relay that performs the swap
faithfully but may misreport its outcome, and asked how often it lies
on the pairs that are delivered and kept.

The noiseless ideal functionality
cannot be realised composably, with or without sacrificed test pairs.
By relabelling his outcome, Charlie influences the delivered state,
which the functionality does not allow an adversary to do. Adding the
honest noise to the ideal does not help, since it is still not realised
at a misreport rate of zero. A useful guarantee must therefore concede
a small misreport rate. 

With this in mind, we proposed a cut-and-choose protocol that 
certifies that rate against an agreed minimum noise level, composably, and
without consuming the pairs it certifies. Unlike purification-based
approaches~\cite{lo1999unconditional,aschauer2002private}, the
guarantee is of finite size and needs no distillation. Unlike mediated key
distribution~\cite{krawec2015mediated,liu2018mediated,sohail2026byzantine},
it delivers entangled pairs rather than a key.

The price to pay for this solution is memory. 
The number of test pairs depends as much on the
minimum noise level as on $f_{\mathrm{target}}$, and the whole batch must be
held until the test set is drawn. Certifying any non-trivial rate needs
a few hundred pairs held at once, and certifying a $1\%$ misreport
rate, the largest we expect fault-tolerant applications to tolerate
(\Cref{ssec:target}), needs $10^5$ to $7\times10^5$.
This demand is out of reach for any demonstrated memory to date in both
capacity and coherence. 
Because the memory adds noise of its own, it
must also last $40$ to $50$ times longer than the batch. 

To ease this, we proposed block-wise verification, which releases
pairs as each block is verified. Memory then no longer limits how many
pairs can be delivered, but it still limits how small a misreport rate
can be certified. A $10^4$-qubit memory, for example, delivers
$4.7\times10^5$ pairs but can only certify that at most $14\%$ of
them were misreported,
and none of the block sizes we considered reaches the $1\%$ target
(\Cref{ssec:blocks}).

\subsection{Implications for the quantum internet}\label{ssec:implications}

When a batch is accepted, our protocol bounds the fraction of delivered
pairs that Charlie misreported by $f_{\mathrm{target}}$, counting only
errors above the minimum noise levels $\gamma_Z, \gamma_X$ agreed in
advance. Since a misreport cannot be told apart from readout noise
(\Cref{prop:unidentifiable}), this bound can only ever be stated
relative to those agreed levels. 
In a network this can work like a
quality-of-service agreement. The provider that operates the relay declares $\gamma_Z$ and
$\gamma_X$, the users decide whether that noise plus
$f_{\mathrm{target}}$ is good enough for their application, and the
protocol holds the provider to it. 
A rejected batch can show that the pairs
were worse than agreed, but not whether Charlie lied or his hardware
failed.
Link-layer protocols proposed for quantum networks already let a
request specify a minimum fidelity, which the link either meets with high confidence or rejects
as unachievable~\cite{dahlberg2019link}. Our protocol would enable the future users of the quantum internet to 
hold a relay run by another provider to such a commitment.

A provider could, however, declare noise levels higher than the true
ones. Charlie could then misreport within the difference without being
caught, and a level declared $\Delta$ too high lets him misreport an
extra fraction $\Delta/(1-\gamma_{\max})$ of the pairs
(\cref{eq:inflated}). The users could see the declared levels before the
batch starts and could refuse them. 
Inflated levels also cannot make the pairs worse than the users agreed
to, since \Cref{prop:fidelity} bounds the total error of an accepted
batch by the declared levels plus a small sampling margin.
Beyond that, the only way to reduce how much Charlie can misreport
undetected is to agree lower levels, which needs Bell measurements with
less honest noise. 
Less noisy measurements would also lower the cost of the test, since
fewer test pairs are needed at a lower level (\cref{eq:k-scaling}), and
a lower channel error rate $\delta$ lets a looser $f_{\mathrm{target}}$
fit within the same error-correction budget (\Cref{ssec:target}). Better Bell measurements improve security as well
as fidelity.

The verification in this work rests on the users preparing the pairs
themselves, so that they know the state Charlie receives
(\Cref{sec:es}). A relay that is also the source,
as in the standard architecture, leaves the users nothing to compare
against unless they inject reference values of their own, as in
mediated key distribution~\cite{krawec2015mediated,liu2018mediated}.
This means third-party nodes can be held to account only in
architectures where they just measure, such as heralded midpoint
links~\cite{pompili2021realization,stolk2024metropolitan}, which
networks should weigh when deciding where to place their sources.

Nonetheless, until end nodes can hold $10^5$ to $7\times10^5$ pairs for days to weeks
(\Cref{tab:joint}), networks built from third-party relays will have to
trust their reports, or restrict themselves to applications such as
MDI-QKD~\cite{lo2012measurement}, which tolerates an untrusted relay
because it needs a key rather than entanglement. Repeater networks
already need long-lived memories~\cite{wehner2018quantum}, and security
adds a requirement of its own on top. With smaller memories, the bounds
that block-wise verification reaches (\Cref{ssec:blocks}) are too loose
for fault-tolerant applications but enough to catch a relay that is
badly faulty or misbehaving. At that scale the protocol is better suited
to network management, flagging relays to recalibrate or avoid, than to
security in the strict sense.

\subsection{Open questions}\label{ssec:open}

Three questions remain open. 
1) First, the memory cost must come down
before the protocol can run on real hardware, whether through sharper
sampling bounds or by not holding the whole batch until the test set is
drawn. 
2) Second, our threat model excludes a relay that keeps side
information about the delivered pairs. Distillation and
device-independent tests can remove it, but they consume the pairs, and
we do not know whether it can be bounded without doing so. 
3) Third, a
repeater chain has more than one relay. Splitting $f_{\mathrm{target}}$
and $\varepsilon_{\mathrm{sec}}$ across hops is straightforward, but
the memory need grows with each hop, so on a quantum internet the number
of third-party relays on a path becomes a cost that routing should take
into account.

Lies that look like noise can
therefore be bounded, but only by holding many pairs at once, and it is
quantum memory that will decide when this
becomes practical.

\ifanon\else
\begin{acknowledgments}
The authors are grateful to Alex Cojocaru for insightful discussions and pointers on this subject.
This research was supported by the EPSRC UK Quantum Technologies Programme under grant EP/T001062/1 and VeriQloud.
\end{acknowledgments}
\fi

\section*{Use of AI tools}
The large language model Claude Opus 5.5 (Anthropic) was used in
preparing this paper, to assist with writing and revising the
manuscript and the accompanying code. The authors reviewed all
AI-assisted content and take full responsibility for the paper.

\section*{Data availability}
The code that generates the numerical results and tables of this paper
is openly available at
\ifanon
\url{https://anonymous.4open.science/r/Untrusted-relay-entanglement-swap-8AE6/}.
\else
\url{https://github.com/grageragarces/Untrusted-relay-entanglement-swap},
and archived at \url{https://doi.org/10.5281/zenodo.24372164}.
\fi

\appendix

\FloatBarrier

\section*{Appendices}

\section{Notation}\label{app:notation}

\Cref{tab:notation} collects the notation used throughout this paper.

\begin{table*}[t]
\caption{Notation, grouped by where it is introduced.}\label{tab:notation}
\begin{ruledtabular}
\begin{tabular}{ll}
$A_1,A_2,B_1,B_2$ & Alice's and Bob's qubit halves; $A_2,B_2$ sent to
  Charlie, $A_1,B_1$ retained \\
$r$ & Charlie's true Bell-measurement outcome \\
$m$ & Charlie's announced report ($m^{(i)}$ on pair $i$); $m=r$ if honest \\
$\rho_{AB}$, $\rho^{\mathrm{hon}}_{AB}$ & the delivered state; its
  honest-execution counterpart \\
$F(\cdot,\cdot)$ & fidelity \\
$C^*$ & a dishonest Charlie \\
$e$, $P_e$ & Charlie's deviation label and the Pauli it induces,
  \Cref{lem:attack} \\
\hline
$\mathcal{F}_0$ & the noiseless ES functionality, \Cref{def:ideal-F} \\
$\mathcal{F}^{M}_{\Lambda,f}$ & the relaxed ES functionality, \Cref{def:F-relaxed} \\
$\mathcal{E}$ & an environment, \Cref{def:environment} \\
$S$ & a simulator, \Cref{def:simulator} \\
$\eta$ & composable security error, \Cref{def:composable} \\
$\mu$ & completeness parameter, \Cref{def:completeness} \\
$q$, $p_B$ & the least Bell-diagonal weight; the weights themselves,
  \Cref{thm:nogo} \\
\hline
$N = M+k$ & pairs prepared; $M$ delivered, $k$ tested \\
$n = k/2$ & pairs tested in each basis \\
$T = T_Z \sqcup T_X$ & the test set and its balanced partition \\
$D$ & the delivered set, $\{1,\dots,N\}\setminus T$ \\
$\pi$ & the delivery permutation of $D$, \Cref{rem:symmetrisation} \\
$\gamma_Z, \gamma_X$ & honest noise baselines; $\gamma_{\max}$ the largest per-pair baseline (\Cref{rem:index-noise}) \\
$\delta$ & infidelity of the honest delivered state \\
$\Lambda$ & the honest channel on a delivered pair \\
$f$ & misreport rate on the delivered pairs, \cref{eq:f-def} \\
$f_N$ & the same rate over the whole batch \\
$f_b$ & rate visible in basis $b$; $f_x, f_z, f_y$ by Pauli \\
$f_{\mathrm{target}}$ & the tolerated bound on $f$ \\
$a_b$ & anti-correlated fraction in basis $b$; $\hat{a}_b$ observed \\
$\phi_b, \phi^D_b$ & density of basis-$b$-visible deviations \\
\hline
$\theta$ & acceptance margin \\
$\nu$ & estimation margin \\
$t$ & sampling tolerance \\
$\varepsilon_{\mathrm{sec}}$ & failure probability; $L = \log(6/\varepsilon_{\mathrm{sec}})$ \\
\hline
$\tau_{\mathrm{swap}}$ & time per delivered swap round \\
$T_{\mathrm{coh}}$ & required coherence time, \cref{eq:coherence} \\
$N_{1/e}$ & memory decay constant, in entangling
  attempts~\cite{pompili2021realization} \\
\hline
$B$ & number of blocks; $n_b = N/B$ the block size,
  Appendix~\ref{app:blocks} \\
$k_b$ & test pairs drawn within each block, $k = Bk_b$ \\
$T_j$ & the test set drawn within block $j$ \\
$\bar{a}_{b,j}$ & mean anti-correlation rate over block $j$ in basis $b$;
  $\bar{a}_b$ its average over the blocks \\
$\hat{S}$ & pooled anti-correlated count over the $k/2$ pairs tested in
  basis $b$, \Cref{lem:pooled} \\

\end{tabular}
\end{ruledtabular}
\end{table*}

\section{Security of quantum communication primitives} \label{app:related-work}

This appendix gives the reasoning behind each row of \Cref{tab:related}.
Only the MDI and mediated key-distribution protocols take a classical
report about an unrepeatable measurement as the object of suspicion,
and only purification-based QKD, private entanglement via repeaters,
quantum authentication, and pure-state and Serfling-based verification
keep some of what they certify.

\begin{itemize}
\item \emph{Self-testing}~\cite{fu2022parallel} certifies, statistically
and noise-tolerantly, that an untrusted device prepared $N$ EPR pairs;
the tested pairs are consumed, and the setting is a single device.
\item \emph{Device-independent testing}~\cite{coladangelo2016parallel}
certifies from a Bell violation that two non-communicating, untrusted
provers hold a given state; our untrusted party is the intermediary.
\item \emph{Quantum authentication}~\cite{barnum2002authentication}
protects a state in transit under a secret key, not classical data
reported after a measurement, and consumes auxiliary qubits.
\item \emph{Graph-state verification}~\cite{unnikrishnan2022verification}
extends entanglement verification to noise and to multipartite resource
states, and consumes what it certifies. \emph{Pure-state
verification}~\cite{zhu2019efficient,zhu2019general} gives tight sample
complexities against an adversarial source, keeps one untested copy,
and accepts only if every test passes. \emph{Serfling-based
verification}~\cite{takeuchi2019resource} tolerates some failed tests
and can keep several copies, a vanishing fraction of those prepared,
with a guarantee on their joint fidelity. All three certify what a
source emits.
\item \emph{Purification-based QKD}~\cite{lo1999unconditional,shor2000simple}
and \emph{private entanglement via
repeaters}~\cite{aschauer2002private} test a random subset of pairs
delivered through repeaters under an adversary's control, and distil
the rest into pairs whose correlations with the adversary vanish. Their
adversary is strictly stronger than ours and covers misreporting
implicitly, but the guarantee concerns the distilled state, the
analysis is asymptotic, and distillation needs bilateral two-qubit
operations or fault-tolerant computation.
\item \emph{MDI-QKD}~\cite{lo2012measurement} is architecturally the
closest, with an untrusted relay performing a Bell measurement, and
shows the architecture is compatible with composable security given a
suitable ideal functionality; its finite-key
analysis~\cite{curty2014finite} postpones the choice of test signals
until after the relay's announcement, as (A3) does. Its deliverable is a
classical key.
\item \emph{Mediated semi-quantum key
distribution}~\cite{krawec2015mediated,liu2018mediated} has an
untrusted server perform Bell measurements and report the results, and
checks the reports against values fixed by the users' own choices;
Liu and Hwang analyse fake Bell-measurement announcements explicitly,
over noiseless channels. \emph{Misreport traps in mediated
conference-key agreement}~\cite{sohail2026byzantine} detect repeaters
that misreport their announcements, against a calibrated honest
baseline and at finite size. In all three the deliverable is a key and
the check decides whether to abort; none bounds the rate of misreports
on pairs that are kept.
\emph{MDI entanglement witnesses}~\cite{branciard2013mdiew} carry the
idea to entanglement certification, suspecting the device's classical
output as we do, but return a witness value rather than usable pairs.
\item \emph{Bell-measurement
self-testing}~\cite{renou2018selftesting,supic2020selftesting} certifies
that Charlie's measurement is the prescribed one, which
\Cref{ssec:threat-scope} assumes, and says nothing about the report
announced afterwards, which is what we test.
\item \emph{Bilocality tests}~\cite{branciard2010characterizing,branciard2012bilocal}
certify that a swap between independent sources genuinely occurred, with
an uncharacterised intermediate node as here; the deliverable is a
certificate about correlations.
\item \emph{Verifiable blind computation}~\cite{fitzsimons2017unconditionally,gheorghiu2019verification}
hides trap qubits among computation qubits, the same random-subset
argument we use, under noise; the traps are consumed and the object
under test is a quantum operation.
\item \emph{Multipartite entanglement
verification}~\cite{pappa2012multipartite} is the closest existing
structure: a random test a dishonest subset of parties cannot
anticipate, analysed in the worst case. The untrusted party holds a
share of the state rather than performing the swap, and the tested
state is consumed.
\end{itemize}

\section{Derivation of the security bound}\label{app:security-analysis}

This appendix gives the derivations outlined in
\Cref{sssec:outline}, with $L$, $n = k/2$, $g$, $\theta$, $\nu$ and $t$
as in~\cref{eq:budget}.

\subsection{False rejection} \label{sssec:false-rejection}

We first show that an honest execution is rarely rejected. It is
rejected if, in either basis, the observed anti-correlated fraction
exceeds its threshold:
\begin{equation}
  \Pr(\bot \mid \text{honest}) \leq
  \sum_{b \in \{Z,X\}}
  \Pr\!\big(\hat{a}_b > \tfrac{\gamma_b}{2} + \theta\big).
\end{equation}
Fix a basis $b$. In an honest execution the average anti-correlation
probability is $\bar{p} = \gamma_b/2$, so \Cref{lem:conc} with
$n = k/2$ gives
\begin{equation}
  \Pr\!\big(\hat{a}_b > \tfrac{\gamma_b}{2} + \theta\big)
  \leq \exp\!\left(-\frac{n\theta^2}
    {2\tfrac{\gamma_b}{2}(1-\tfrac{\gamma_b}{2}) + \tfrac{2}{3}\theta}\right).
\end{equation}
The threshold $\theta$ in~\cref{eq:budget} is chosen so that this bound
equals $\varepsilon_{\mathrm{sec}}/6$ at the larger baseline
$\gamma_{\max}$. Since the bound grows with the baseline, it is at most
$\varepsilon_{\mathrm{sec}}/6$ in both bases, and
\begin{equation}\label{eq:false-reject}
  \Pr(\bot \mid \text{honest}) \leq \varepsilon_{\mathrm{sec}}/3 .
\end{equation}
The test is one-sided: an $\hat{a}_b$ below the baseline only means the
hardware is performing better than declared, which no deviation of
\Cref{lem:attack} can cause.

\subsection{False acceptance} \label{sssec:false-acceptance}

We first bound the batch rate $f_N$ on acceptance, then transfer the
bound to the delivered rate $f$.
Suppose the true rate satisfied
$a_b > \gamma_b/2 + \theta + \nu$. Applying the lower tail
of~\cref{eq:conc} in basis $b$ at deviation
$s = a_b - \gamma_b/2 - \theta > \nu$, and using
$a_b(1-a_b) \leq a_b = \gamma_b/2 + \theta + s$ to remove the unknown
rate from the variance, the probability of nonetheless observing
$\hat{a}_b \leq \gamma_b/2 + \theta$ is at most
\begin{equation}
  \exp\!\left(-\frac{n s^2}
  {2\bigl(\tfrac{\gamma_b}{2}+\theta\bigr) + \tfrac{8}{3}s}\right)
  \;\leq\; \varepsilon_{\mathrm{sec}}/6 ,
\end{equation}
the last step holding for every $s \geq \nu$ because $\nu$
is the positive root of the quadratic that turns it into an equality.
Over both bases this is $\varepsilon_{\mathrm{sec}}/3$. We work from here on the complementary event, on which
\begin{equation}
  a_b - \tfrac{\gamma_b}{2} \;\leq\; \theta + \nu
\end{equation}
in both bases. \Cref{lem:visibility} bounds the same quantity from
below by the worst case over Charlie's choice of deviation:
\begin{equation}\label{eq:lowerbound}
  \max_{b} \Bigl(a_b - \tfrac{\gamma_b}{2}\Bigr)
  \geq \frac{f_N}{2}\bigl(1-\gamma_{\max}\bigr).
\end{equation}
Combining the two,
\begin{equation}\label{eq:fN-bound}
  f_N \;\leq\; \frac{2(\theta + \nu)}{1-\gamma_{\max}}.
\end{equation}

It remains to convert~\cref{eq:fN-bound} into a statement about the $M$
delivered pairs. This is the cut-and-choose step, in the form used for
QKD parameter estimation~\cite{bouman2010sampling,tomamichel2012tight}.
Fix a basis $b$ and let $V_b \subseteq \{1,\dots,N\}$ be the set of
pairs whose deviation is visible in basis $b$; write $\phi_b = |V_b|/N$ for its density in the batch and
$\phi_b^D$ for its density among the $M$ delivered pairs. By (A3), membership of $V_b$ is fixed before $T$ is drawn, so
$|V_b \cap T|/k$ is the mean of a sample of size $k$ drawn without
replacement from a population of $N$ pairs, a fraction $\phi_b$ of
which lie in $V_b$. The standard concentration bound for such
samples~\cite{serfling1974probability} gives
\begin{equation}\label{eq:hypergeom}
  \Pr\Bigl[\, |V_b \cap T|/k \;\leq\; \phi_b - t \,\Bigr]
  \;\leq\; \exp\!\left(-\frac{2kt^2}{1-(k-1)/N}\right),
\end{equation}
which is $\varepsilon_{\mathrm{sec}}/6$ per basis, hence $\varepsilon_{\mathrm{sec}}/3$ over both,
by~\cref{eq:budget}. $|V_b \cap T|$ is never observed; only
the lower bound is needed, and it holds whatever Charlie's deviation
pattern is. The finite-population factor $1-(k-1)/N$ stops the transfer
term from growing without limit as $k$ approaches $N$; discarding it
roughly doubles the smallest $f_{\mathrm{target}}$ the analysis can
certify (\Cref{ssec:test-overhead}).

The delivered pairs are the untested ones, so
$N\phi_b = |V_b \cap T| + M\phi_b^D$. On the complement
of~\cref{eq:hypergeom},
\begin{equation}\label{eq:transfer}
  M\phi_b^D \;=\; N\phi_b - |V_b \cap T|
  \;\leq\; N\phi_b - k(\phi_b - t) \;=\; M\phi_b + kt,
\end{equation}
that is, $\phi^D_b \leq \phi_b + (k/M)\,t$, an \emph{additive}
correction. We work with the densities $\phi_b$ rather than with
anti-correlation rates on the delivered pairs, whose honest baseline need
not equal $\gamma_b$ (\Cref{rem:index-noise}).

Two facts turn~\cref{eq:transfer} into the guarantee. First,
\cref{eq:per-basis-a} reads $a_b - \gamma_b/2 = \phi_b(1-\gamma_b)$, so
the upper bound $a_b - \gamma_b/2 \leq \theta + \nu$ established
above gives
\begin{equation}\label{eq:phi-bound}
  \phi_b \;\leq\; \frac{\theta+\nu}{1-\gamma_b}
  \;\leq\; \frac{\theta+\nu}{1-\gamma_{\max}} .
\end{equation}
Second, the proof of \Cref{lem:visibility} establishes
$\phi_Z + \phi_X = f_N + f_y \geq f_N$ by counting alone, so the same
step holds on any subset of the batch fixed before $T$ is drawn; applied
to the delivered set it gives $\max_b \phi^D_b \geq f/2$. Chaining the
three,
\begin{equation}\label{eq:f-chain}
  \frac{f}{2} \;\leq\; \max_b \phi^D_b
  \;\leq\; \max_b \phi_b + \frac{k}{M}\,t
  \;\leq\; \frac{\theta+\nu}{1-\gamma_{\max}} + \frac{k}{M}\,t .
\end{equation}
Relaxing the last term by $1 \leq (1-\gamma_{\max})^{-1}$
gives~\cref{eq:f-final}, so every $k$ reported in
\Cref{sec:resources}, which solves the relaxed condition, is
conservative for~\cref{eq:f-chain}.
Requiring $f \leq f_{\mathrm{target}}$ therefore amounts to
\begin{equation}\label{eq:n-threshold}
  2(\theta + \nu) + \frac{2k}{M}\,t
  \;\leq\; f_{\mathrm{target}}\,(1-\gamma_{\max}).
\end{equation}

\section{Block-wise verification}\label{app:blocks}
Running the batch in $B$ blocks of $n_b = N/B$ pairs, with each block's
outcomes broadcast, its test subset drawn, and the block verified before
the next begins, removes the need to hold a pair through the whole
batch. The obstacle to treating the blocks together is that Charlie
sees the test sets and outcomes of blocks $1,\dots,j-1$ before
committing to block $j$, so his deviation densities are adaptive and
\Cref{lem:conc} does not apply to the pooled sample directly.

\begin{lemma}[Pooled concentration under adaptive blocking]\label{lem:pooled}
Partition the batch into $B$ blocks of $n_b$ pairs. Suppose that for
each $j$ Charlie fixes his deviation pattern on block $j$ as an
arbitrary function of everything revealed in blocks $1,\dots,j-1$, and
that only afterwards is the test set $T_j$ drawn within block $j$,
uniformly and with a balanced partition, as in step~3. Fix a basis $b$,
write $\bar{a}_{b,j}$ for the mean anti-correlation rate over block $j$
and $\bar{a}_b = B^{-1}\sum_{j}\bar{a}_{b,j}$, and let $\hat{S}$ be the
total anti-correlated count over the $k/2$ pairs tested in basis $b$
across all blocks, $k_b/2$ from each.
Then for every $\lambda \in \mathbb{R}$,
\begin{equation}\label{eq:pooled-mgf}
  \mathbb{E}\Bigl[\exp\bigl(\lambda\hat{S}
    - \tfrac{k}{2}\log(1-\bar{a}_b+\bar{a}_b e^{\lambda})\bigr)\Bigr]
  \;\leq\; 1 .
\end{equation}
Consequently:
\begin{enumerate}
\item[(i)] if Charlie is honest, so that $\bar{a}_b$ is fixed in
advance, every Chernoff bound available for $k/2$ independent
$\mathrm{Bernoulli}(\bar{a}_b)$ draws holds for $2\hat{S}/k$, in both
tails;
\item[(ii)] for any constants $0 \leq c < q \leq 1$ fixed in advance,
\begin{equation}\label{eq:pooled-threshold}
\begin{split}
  \Pr\Bigl[\tfrac{2}{k}\hat{S} \leq c \;\wedge\; \bar{a}_b \geq q\Bigr]
  &\leq e^{-\frac{k}{2} D(c\,\|\,q)} \\
  &\leq \exp\!\left(-\frac{k}{2}\,
    \frac{(q-c)^2}{2q + \tfrac{2}{3}(q-c)}\right),
\end{split}
\end{equation}
with $D(c\,\|\,q)$ the binary relative entropy.
\end{enumerate}
These are the two uses the analysis makes of it: $\theta$ concerns an
honest execution, and $\nu$ bounds an event of the form (ii) with
$c = \gamma_b/2 + \theta$ and $q = c + \nu$.
\end{lemma}
\begin{proof}
Fix $b$ and $\lambda$, write $h(x) = \log(1-x+xe^{\lambda})$, and let
$M_j$ be the anti-correlated count on $T_{j,b}$. Let $\mathcal{F}_j$ be
generated by everything revealed through the end of block $j$ together
with Charlie's private randomness. By hypothesis his pattern on block
$j$, and hence $\bar{a}_{b,j}$, is $\mathcal{F}_{j-1}$-measurable, while
$T_{j,b}$ is drawn uniformly within block $j$ afterwards and the
outcomes on it are conditionally independent given the rates. Applying
\Cref{lem:conc} within block $j$, conditionally on $\mathcal{F}_{j-1}$,
\begin{equation}
  \mathbb{E}\bigl[e^{\lambda M_j} \mid \mathcal{F}_{j-1}\bigr]
  \;\leq\; e^{\,(k_b/2)\,h(\bar{a}_{b,j})} .
\end{equation}
Put $Z_0 = 1$ and
$Z_j = Z_{j-1}\exp(\lambda M_j - (k_b/2)\,h(\bar{a}_{b,j}))$.
The display states that
$\mathbb{E}[Z_j \mid \mathcal{F}_{j-1}] \leq Z_{j-1}$, so $(Z_j)$ is a
supermartingale and $\mathbb{E}[Z_B] \leq 1$, which is
\begin{equation}\label{eq:pooled-super}
  \mathbb{E}\Bigl[\exp\Bigl(\lambda\hat{S}
    - \tfrac{k_b}{2}\sum_{j=1}^{B} h(\bar{a}_{b,j})\Bigr)\Bigr] \;\leq\; 1 .
\end{equation}
Finally $h''(x) = -(e^{\lambda}-1)^2/(1+x(e^{\lambda}-1))^2 \leq 0$, so
$h$ is concave and Jensen's inequality gives
$\sum_j h(\bar{a}_{b,j}) \leq B\,h(\bar{a}_b)$ pathwise. Substituting
this into~\cref{eq:pooled-super} gives~\cref{eq:pooled-mgf}, since
$Bk_b = k$.

For (i), $\bar{a}_b$ is deterministic, and Markov's inequality applied
to~\cref{eq:pooled-mgf} at each fixed $\lambda$ is the Chernoff argument
for $\mathrm{Binomial}(k/2,\bar{a}_b)$. For (ii), take $\lambda < 0$.
Then $h$ is decreasing, so on the event in~\cref{eq:pooled-threshold}
we have $\lambda\hat{S} \geq \lambda\tfrac{k}{2}c$ and
$h(\bar{a}_b) \leq h(q)$, and the exponent in~\cref{eq:pooled-mgf} is
at least the constant $\tfrac{k}{2}(\lambda c - h(q))$. Markov's
inequality bounds the probability of the event by
$\exp(-\tfrac{k}{2}(\lambda c - h(q)))$, and optimising over
$\lambda < 0$ gives $D(c\,\|\,q)$. The Bernstein form follows because it
is derived from the same moment generating function, with the variance
$q(1-q)$ bounded by $q$.

The restriction to fixed thresholds matters. With $\bar{a}_b$ random, a
single $\lambda$ yields only a bound linear in $\bar{a}_b$; a tail
bound of Bernstein shape in the \emph{realised} $\bar{a}_b$ would need
a union over a grid of $\lambda$ or a mixture over
$\lambda$~\cite{howard2021time}. The analysis never needs one.
\end{proof}

Adaptivity therefore does not weaken the bound: what Charlie learns
between blocks changes $\bar{a}_{b,j}$, but cannot move the conditional
law of $M_j$ away from the block mean, because $T_j$ is still drawn
after he has committed. Pooling leaves $\theta$ and $\nu$ unchanged. The sampling
transfer chains the same way: Serfling's bound is itself derived from a
moment generating function inequality that keeps the finite-population
factor~\cite[Thm~2.2]{serfling1974probability}, so applying it within
each block, with the threshold $t$ fixed in advance, and chaining as
above gives~\cref{eq:hypergeom} with $1-(k-1)/N$ replaced by
$1-(k-B)/N$. At the operating point this changes $k$ by at most four
pairs for $B \leq 100$ ($168{,}982$ becomes $168{,}986$), and the floor
on $f_{\mathrm{target}}$ stays at $0.38\%$.
Verifying each block independently against $\varepsilon_{\mathrm{sec}}/B$
would instead cost $14.1$ times the monolithic count at
$f_{\mathrm{target}} = 0.1$ with blocks of $10^5$, and admit no solution
for blocks of $10^4$ or at $f_{\mathrm{target}} = 10^{-2}$.

The release modes this enables, and what each saves, are discussed in
\Cref{ssec:blocks}.

\bibliography{references}

\end{document}